\documentclass[aps,prb,twocolumn,groupedaddress,showpacs,floatfix,altaffilletter,longbibliography]{revtex4-2}
\usepackage{graphicx}
\usepackage{grffile}
\usepackage{amsmath, amssymb, amsfonts}
\usepackage{physics}
\usepackage{bm}
\usepackage{color}
\usepackage[dvipsnames]{xcolor}
\usepackage{epsfig}
\usepackage{times}
\usepackage{hyperref}
\hypersetup{colorlinks,bookmarks=false,citecolor=blue,linkcolor=red,urlcolor=blue}
\usepackage{textcomp, gensymb}
\usepackage[toc,page]{appendix}
\usepackage{placeins}

\newcommand{\figref}[1]{Fig.~\ref{#1}}

\newcommand{\secref}[1]{Sec.~(\ref{#1})}

\newcommand{\eref}[1]{Eq.~(\ref{#1})}

\newcommand{\figwidth}{0.97\columnwidth}

\newcommand{\Gfn}{\mathcal{G}}

\begin{document}

\title{Nonlocal corrections to dynamical mean-field theory from the two-particle self-consistent method}

\author{N. Martin, C. Gauvin-Ndiaye, and A.-M. S. Tremblay}
\affiliation{D\'epartement de physique, Regroupement qu\'eb\'ecois sur les matériaux de pointe \& Institut quantique\\
Universit\'e de Sherbrooke, 2500 Boul. Universit\'e, Sherbrooke, Qu\'ebec J1K2R1, Canada}
\date{\today}
\begin{abstract}

Theoretical methods that are accurate for both short-distance observables and long-wavelength collective modes are still being developed for the Hubbard model. 
Here, we benchmark an approach that combines dynamical mean-field theory (DMFT) observables with the two-particle self-consistent theory (TPSC). This offers a way to include non-local correlations in DMFT while also improving TPSC. The benchmarks are published diagrammatic quantum Monte Carlo results for the two-dimensional square lattice Hubbard model with nearest-neighbor hopping. 
This method (TPSC+DMFT) is relevant for weak to intermediate interaction, satisfies the local Pauli principle and allows us to compute a spin susceptibility that satisfies the Mermin-Wagner theorem. 
The DMFT double occupancy determines the spin and charge vertices through local spin and charge sum rules. 
The TPSC self-energy is also improved by replacing its local part with the local DMFT self-energy.
With this method, we find improvements for both spin and charge fluctuations and for the self-energy. 
We also find that the accuracy check developed for TPSC is a good predictor of deviations from benchmarks for this model. 
TPSC+DMFT can be used in regimes where quantum Monte Carlo is inaccessible. 
In addition, this method paves the way to multi-band generalizations of TPSC that could be used in advanced electronic structure codes that include DMFT.   

\end{abstract}

\maketitle

\section{Introduction}

The Hubbard model is among the simplest models that enable the study of strong electronic correlations. 
Despite its apparent simplicity, exact solutions to this model only exist in one dimension \cite{Lieb_1968} and in the limit of infinite dimensions \cite{Metzner_1989}. 
Numerically exact methods such as quantum Monte Carlo (QMC) and diagrammatic Monte Carlo can also provide, within statistical uncertainty, useful results. 
However, such methods are generally plagued with a sign problem,  can become too computationally expensive for large system sizes, or are difficult to extend to multi-orbital models, which are relevant to the study of realistic systems.

Dynamical mean field theory (DMFT) is one of the most widely used methods that gives accurate results for static and frequency-dependent local quantities over a wide range of interaction strengths, making possible the description of the Mott metal-insulator transition~\cite{Metzner_1989, Georges_1992, Jarrell:1992, georges_dynamical_1996}. 
However, non-local spatial correlations are missing in DMFT, which approximates the self-energy as purely local. 
These non-local correlations are particularly crucial to the correct description of physical phenomena encountered in the weak to intermediate coupling regimes of low dimensional systems.
Extensions of DMFT such as cluster methods \cite{Hettler_1998, Lichtenstein_2000, Kotliar_2001,Maier:2005,KotliarRMP:2006,LTP:2006} and diagrammatic extensions~\cite{Toschi_Katanin_Held_2007,Kusunose_2006,Rubtsov_Katsnelson_Lichtenstein_2008,Rubtsov_Katsnelson_Lichtenstein_2012,Aryal_Parcollet_2015,Aryal_Parcollet_2017,Aryal_Parcollet_2016,Rohringer_2018}  can be used to circumvent this limitation of DMFT. 

Other approaches based on the addition of the non-local part of the self-energy coming from a weak-coupling approach to the local DMFT self-energy have also been proposed. 
These methods include GW+DMFT \cite{Tomczak_2017, Sun_2002, Biermann_2003, Ayral_2012, Ayral_2013},  DMFT+FLEX \cite{Gukelberger_2015,Kitatani_2015}, as well as DMFT+TMA \cite{Gukelberger_2015}. In these methods, the local part of the GW, FLEX or TMA self-energy is replaced by the local DMFT self-energy. In DMFT+$\Sigma_{\mathbf{k}}$ \cite{Sadovskii_2005, Kuchinskii_2005, Kuchinskii_2012}, the DMFT self-energy is simply added. 
More recently, an approximation that combines the non-local part of a self-energy built from the weak-coupling D$\Gamma$A spin susceptibility  with the local DMFT self-energy was found to be qualitatively accurate in the weak-coupling regime of the two-dimensional ($2D$) Hubbard model \cite{Schafer_2021}. 
These approaches can be motivated by the high accuracy of the local self-energy obtained from DMFT, as evidenced by benchmarks in the weak coupling regime \cite{Gukelberger_2015, Schafer_2021, Simkovic_2022}. 
However, benchmarks of methods mixing DMFT and perturbative approaches (GW, FLEX and TMA) against diagrammatic Monte Carlo calculations have shown that these uncontrolled schemes perform poorly in the weakly interacting regime and can yield inaccurate results for the non-local part of the self-energy \cite{Gukelberger_2015}. 
It then becomes relevant to investigate an approach that combines   local DMFT observables with non-local fluctuations contained in a non-perturbative method.

The two-particle self-consistent (TPSC) approach is one such non-perturbative method that was first developed for the one-band Hubbard model~\cite{Vilk:1994,Vilk:1996}. For reviews, see Refs.~\cite{Vilk_1997,Allen:2003,TremblayMancini:2011}.
It is valid in the weak to intermediate coupling regime of the Hubbard model, respects both the Pauli principle and the Mermin-Wagner theorem, and satisfies the local spin and charge sum rules. 
It has since been extended to multi-orbital~\cite{Miyahara_2013, Zantout_2021} and multi-site cases~\cite{Aizawa_2015,Arya_Sriluckshmy_Hassan_Tremblay_2015,Ogura_Kuroki_2015,Zantout_2018,Zantout_2019,Pizarro_Adler_Zantout_Mertz_Barone_Valenti_Sangiovanni_Wehling_2020}.
 TPSC has been benchmarked against DiagMC calculations \cite{Schafer_2021} and finite size Monte Carlo~\cite{Vilk:1994,Veilleux:1995,Vilk:1996,Vilk_1997,Moukouri:2000,Kyung_2001,TremblayMancini:2011}. 
The TPSC approach has the advantage of being computationally inexpensive, which enables its application and extension to a wide array of problems. 

For example, this method has been used to study multiple facets of the $2D$ Hubbard model, such as the weak-coupling pseudogap \cite{Vilk_1997, Kyung_2004}, the optical conductivity \cite{Bergeron_2011}, adiabatic cooling in cold atoms~\cite{Dare:2007}, the interplay of disorder and spin fluctuations \cite{Gauvin-Ndiaye_Graham_2022}, the resilience of Fermi liquid quasiparticles on cold parts of the Fermi surface \cite{Gauvin-Ndiaye_2022}, unconventional superconductivity~\cite{Kyung:2003,Hassan:2008,Ogura_Kuroki_2015,Zantout_2018,Otsuki_TPSC_2012}, ferromagnetism~\cite{Hankevych:2003}, crossover of antiferromagnetic spin fluctuations from two to three dimensions~\cite{Dare:1996}, and magnetic properties of the three-dimensional Hubbard model~\cite{Albinet:2000}.
Moreover, TPSC has been generalized to the extended Hubbard model~\cite{Davoudi:2006,Davoudi:2007,Davoudi:2008}, to the attractive model~\cite{Allen:2001,Kyung_2001} and recently to the non-equilibrium case \cite{Simard_2022}. 
Its multi-orbital formulation has also been used in the context of one-shot DFT+TPSC calculations \cite{Zantout_2018, Zantout_2019, Bhattacharyya_2020}. 
These examples illustrate the flexibility of this approach for the Hubbard model.
Though the TPSC approach fails deep in the renormalized classical regime of the $2D$ Hubbard model, its recent improvements named TPSC+ \cite{Schafer_2021, tpscplus} and TPSC+GG \cite{Schafer_2021, Simard_2022} seem to extend its domain of validity down to lower temperatures.

In this work, we introduce another improvement of the TPSC approach, namely the TPSC+DMFT approach for the one-band Hubbard model.  The spin and charge fluctuations are influenced by DMFT results only through the DMFT double occupancy appearing in the local spin and charge sum rules. 
Inspired by previous methods we also combine the non-local part of the TPSC self-energy with the local DMFT self-energy~\cite{Schafer_2021}. This method has also been proposed in the context of the multi-orbital Kanamori model \cite{Zantout_2022}. 
In Sec. \ref{section:theory}, we describe the DMFT and TPSC approaches. 
We then introduce two aspects of the proposed approach in Sec. \ref{section:methods}, namely the replacement of the TPSC \emph{ansatz} with the DMFT double occupancy to compute spin and charge fluctuations, and the $\Sigma^{\mathrm{loc}}$ approach that combines the local part of the DMFT self-energy with the non-local part of the TPSC self-energy.
We compare our results on the $2D$ square lattice to benchmarks from diagrammatic Monte Carlo calculations in Sec. \ref{section:results}. 
We show that the proposed approach, which is only as computationally expensive as single-site DMFT, is valid in the weak to intermediate interaction strength regime $U/t\lesssim 5$, but fails to capture the Mott-Heisenberg physics in the strong interaction regime $5 < U/t < 8$. 
The main results are then summarized in Sec.~\ref{section:summary_main_results}, where we also discuss the domain of validity of the proposed approaches.


\section{Model and theoretical basis}
\label{section:theory}
After introducing the model, we recall in turn the TPSC and DMFT methods. In the latter case, paramagnetic and antiferromagnetic solutions are possible. 
\subsection{The Hubbard Model}
We study the one-band Hubbard model on a square $2D$ lattice, 
\begin{multline}
H = - t\sum_{\langle ij\rangle,\sigma}(c^{\dagger}_{i,\sigma} c_{j,\sigma} + c^{\dagger}_{j,\sigma} c_{i,\sigma})\\
 + U\sum_{i} n_{i,\uparrow} n_{i, \downarrow} - \mu\sum_{i, \sigma} n_{i,\sigma}.
\end{multline}
The nearest-neighbor hopping $t$ is our unit of energy, $U$ is the Hubbard repulsion on a single site and $\mu$ the chemical potential. The occupation number operators are $n_{i,\sigma}$ at lattice position $i$ for spin $\sigma$ while the corresponding annihilation (creation) operators are $c^{(\dagger)}_{i,\sigma}$. Calculations are done for $U=2$ at half filling ($n=1$) to benchmark with Ref.~\cite{Schafer_2021}, and away from half filling for $0.5 < U < 8$ to benchmark with Ref.~\cite{Simkovic_2022_benchmark}. State-of-the-art versions of CDet~\cite{Rossi_2017,Simkovic_2022_benchmark} and $\Sigma$DDMC~\cite{Simkovic_Kozik_2019,Moutenet_Wu_Ferrero_2018} are used in these DiagMC calculations. 

\subsection{The two-particle self-consistent (TPSC) approach}
The two-particle self-consistent approach is a non-perturbative semi-analytical method that satisfies the Pauli principle, the Mermin-Wagner theorem and conservation laws for the spin and charge susceptibilities. These obey the local spin and charge sum rules~\cite{Vilk:1994}
\begin{align}
\frac{T}{N}\sum_{q} \chi_{\mathrm{sp}}(q) &= n-2\expval{n_{\uparrow} n_{\downarrow}}, \label{eq:sumrule_sp}\\
\frac{T}{N}\sum_{q} \chi_{\mathrm{ch}}(q) &= n+2\expval{n_{\uparrow} n_{\downarrow}} - n^2.  \label{eq:sumrule_ch}
\end{align}
We use the convention $q\equiv (\vb{q},iq_n)$, where $q_n=2n\pi T$ is a bosonic Matsubara frequency, $\vb{q}$ is a wave vector in the first Brillouin zone, and $n$ is the filling. 
The susceptibilities are calculated with renormalized local and static vertices $U_{\mathrm{sp}}$ and $U_{\mathrm{ch}}$, and with the non-interacting charge susceptibility $\chi^{(1)}(q)$:
\begin{align}
\chi_{\mathrm{sp}}(q) &\equiv \frac{\chi^{(1)}(q)}{1-\frac{1}{2}U_{\mathrm{sp}}\chi^{(1)}(q)}, \label{eq:chi_sp}\\
\chi_{\mathrm{ch}}(q) &\equiv \frac{\chi^{(1)}(q)}{1+\frac{1}{2}U_{\mathrm{ch}}\chi^{(1)}(q)}. \label{eq:chi_ch}
\end{align}
While the susceptibilities are identical to the non-interacting case, the notation $\chi^{(1)}(q)$ has a conceptual importance to remind us that the self-energy entering the Green's function $\Gfn^{(1)}_{\sigma}(k)$ at this level is a constant that can be absorbed in the chemical potential in such a way that the density is the desired one~\cite{Allen:2003,TremblayMancini:2011}. 

To compute self-consistently the vertices from the sum rules, a third equation is needed. In TPSC, the \emph{ansatz}~\cite{Vilk:1994,Hedayati:1989} on the double occupancy
\begin{equation}
\expval{n_{\uparrow} n_{\downarrow}} =  \frac{U_{\mathrm{sp}}}{U} \expval{n_{\uparrow}} \expval{n_{\downarrow}} \label{eq:ansatz_tpsc}
\end{equation}
is used. It can be seen as a correction to the Hartree-Fock factorization of the double occupancy. More detailed justification is given in Ref.~\cite{Vilk_1997}. In the case of electron-doping, the particle-hole transformed \emph{ansatz} with $n_\sigma \rightarrow 1-n_\sigma$ in \eref{eq:ansatz_tpsc} must be used.

In the spirit of the electron gas, the self-energy is computed in a second step, using the spin and charge susceptibilities that contain collective modes that were obtained while satisfying conservation laws. The formula ~\cite{Vilk:1996,Vilk_1997} that takes into account crossing symmetry~\cite{Moukouri:2000,TremblayMancini:2011} is
\begin{multline}
\Sigma^{(2)}_{\sigma}(k) = Un_{\bar{\sigma}} \\
+\frac{U}{8}\frac{T}{N}\sum_{q}\big[ 3U_{\mathrm{sp}}\chi_{\mathrm{sp}}(q) + U_{\mathrm{ch}}\chi_{\mathrm{ch}}(q) \big] \Gfn^{(1)}_{\sigma}(k+q), \label{eq:self-energie_tpsc}
\end{multline}
where $k=(\vb{k}, ik_n)$, $ik_n=(2n+1)\pi T$ is a fermionic Matsubara frequency, and $\bar{\sigma}$ is the opposite spin.
The best approximation to the Green's function  is $(\Gfn^{(2)})^{-1} = (\Gfn^{(1)})^{-1} - \Sigma^{(2)}$, which contains the second-step self-energy. The chemical potential must be recalculated at this level to keep the filling $n$  constant. 

The spin susceptibility allows us to define a spin correlation length. Around the maximum in the Brillouin zone, located at $\vb{Q}=(\pi, \pi)$ (at half filling), it can be fitted to the Ornstein-Zernicke form
\begin{equation}
\chi_{\mathrm{sp}}(\vb{q},iq_n=0) \approx \frac{A}{(\vb{q}-\vb{Q})^2 + \xi_{\mathrm{sp}}^{-2}}, \label{eq:fit_lorentzienne_xisp}
\end{equation}
with a width at half maximum $\xi_{\mathrm{sp}}^{-1}$ in $\vb{q}$ space. This corresponds to the physical spin correlation length $\xi_{\mathrm{sp}}$.

The accuracy of the TPSC method can be estimated by considering the sum rule
\begin{equation}
    \frac{T}{N}\sum_{k} \Sigma^{(2)}_{\sigma}(k) \Gfn^{(1)}_{\sigma}(k) = U\expval{n_{\uparrow} n_{\downarrow}},
\end{equation}
which, in TPSC, is exactly satisfied with the above definition of the self-energy and double occupancy obtained from the \emph{ansatz} and the spin and charge sum rules. The sum rule on $\Sigma^{(2)}\Gfn^{(2)}$ does not give the above double occupancy. However, it has been observed that the relative violation of the sum rule
\begin{equation}
    \text{violation of sum rule} = \frac{\Tr(\Sigma^{(2)} \Gfn^{(2)}) - U\expval{n_{\uparrow} n_{\downarrow}} } {U\expval{n_{\uparrow} n_{\downarrow}}}, \label{eq:accuracy_check_Sig2G2}
\end{equation}
with $\expval{n_{\uparrow} n_{\downarrow}}$ obtained from the spin and charge sum rules, is small (of the order of a few percents) in the range of parameters where TPSC agrees with exact Monte-Carlo results~\cite{Vilk_1997,TremblayMancini:2011}. This metric can therefore be used to assess the reliability of the TPSC calculation~\footnote{In practice the Hartree contribution is considered separately for convergence purposes.}. Its general usefulness will become apparent in Sec.~\ref{section:discussion_accuracy_check}.

The numerical implementation uses fast-Fourier transforms and the sparse-ir library for Matsubara sums~\cite{Shinaoka_2017, Li_2020, Shinaoka_sparse_2022}.

\subsection{Dynamical mean-field theory (DMFT)}
DMFT yields a finite Néel temperature, hence the question arises whether the paramagnetic (PM) or the antiferromagnetic (AFM) versions of DMFT should be used. It has also been noticed that better results for double occupancy could be obtained in the low temperature regime by allowing for long-range order~\cite{Fratino_AFM:2017}. 
Hence, in this section, we recall a few relevant ideas from dynamical mean-field theory for both the paramagnetic and antiferromagnetic states.
The numerical implementation of both approaches uses the CT-HYB Monte-Carlo impurity solver~\cite{WernerMillis:2006,Gull:2011} implemented in the TRIQS library~\cite{triqs, Seth_cthyb_2016}.

\subsubsection{Paramagnetic DMFT}
In single-site dynamical mean-field theory (DMFT)~\cite{Georges_1992,Jarrell:1992,georges_dynamical_1996}, the fully-interacting lattice problem is mapped onto the Anderson impurity problem, where a single impurity site (treated exactly) is coupled to a non-interacting bath of electrons. The coupling is through a hybridization function, which plays the role of a frequency-dependent mean-field.

The self-consistency between the impurity problem and the lattice problem is enforced by approximating the lattice self-energy as local and equal to the impurity self-energy
\begin{equation}
\Sigma^{\mathrm{lat}}_\sigma(k) = \Sigma^{\mathrm{lat}}_\sigma(\vb{k}, ik_n) \approx \Sigma^{\mathrm{imp}}_{\sigma}(ik_n).
\end{equation}
This approximation becomes exact in the limit of infinite lattice coordination~\cite{Metzner_1989,Georges_1992}. 
The self-consistent self-energy, Green's function, and corresponding observables are computed by an iterative scheme.

\subsubsection{Antiferromagnetic DMFT}
As discussed in Ref.~\cite{georges_dynamical_1996}, DMFT can be extended to describe an antiferromagnetically ordered state (for $n=1$), by considering two sublattices $A,B$ on the square lattice.
For each sublattice, one impurity is isolated.
At each iteration, the impurity problem is solved twice, and self-consistency is imposed on both Green's functions.
The self-consistency equation couples both sublattices together.

To accelerate the convergence to paramagnetic or antiferromagnetic solutions, the self-energy is partially symmetrized at each iteration with
\begin{equation}
\Sigma_{\alpha,\sigma}(ik_n) = \frac{1}{2} \big[ \Sigma_{\alpha,\sigma}(ik_n) + \Sigma_{\bar\alpha,\bar\sigma}(ik_n) \big].
\end{equation}
Here, $\alpha$ stands for the sublattice and a bar over an index represents the opposite value of the index.

An alternative way of allowing an antiferromagnetically ordered solution to stabilize in DMFT is to solve a single impurity problem that we set to be located on sublattice A.
Then, the self-energy for the impurity on sublattice B is set to
\begin{equation}
    \Sigma_{B,\sigma}(ik_n) = \Sigma_{A,\bar{\sigma}}(ik_n).
\end{equation}
This is equivalent to the case where we solve two impurity problems if the number of Monte-Carlo measurements in the CT-HYB impurity solver is doubled (to obtain the same overall statistics).


\section{Proposed approach: TPSC+DMFT}
\label{section:methods}

As proposed earlier~\cite{Vilk_1997,Kyung:2003}, when local quantities such as the double occupancy are available from reliable methods, it is possible to avoid using the TPSC \emph{ansatz}~\eref{eq:ansatz_tpsc} that can be problematic deep in the renormalized classical regime. DMFT offers this possibility. This is discussed in \secref{section:methods_tpscAnsatz}.  In addition, the local self-energy obtained from DMFT is quite accurate, leading to the possibility of combining it with the TPSC self-energy in the way discussed in \secref{section:methods_localSelfEnergy}. 

The widespread availability of DMFT in electronic structure codes also opens the possibility of generalizing TPSC to multiband problems where generalized \emph{ansatz} have not been benchmarked.

\subsection{Replacing the TPSC ansatz}
\label{section:methods_tpscAnsatz}

The first step of a TPSC+DMFT calculation is to insert the DMFT double occupancy in the sum rules Eqs.~\ref{eq:sumrule_sp} and \ref{eq:sumrule_ch} to determine the values of the renormalized vertices. 
We will refer to this method in the next sections as TPSC+DMFT. 
Since the sum rules for the spin and charge susceptibilities are still satisfied exactly, the Mermin-Wagner theorem for $\chi_{\mathrm{sp}}$ and the Pauli principle are still respected by that hybrid method.

In all the DMFT calculations performed in this work, the electronic density is converged within $\pm 0.003$ of the displayed density.
The chemical potential in the TPSC calculation is then chosen to get the same density, to ten significant digits.
As a result, the TPSC+DMFT results shown in Sec.~\ref{section:results} have slightly different electronic densities ($\pm 0.006$) from each other, which results in effective error bars on the TPSC+DMFT results.

\subsection{Correcting the local self-energy}
\label{section:methods_localSelfEnergy}
In the second step, the local self-energy obtained from the TPSC calculation is replaced by the DMFT value \footnote{If the local self-energy is obtained through a AFM-DMFT calculation, the two sublattice self-energies are averaged to get a single AFM-DMFT self-energy.}. As shown in Ref.~\cite{Schafer_2021}, the DMFT local self-energy is more accurate than the corresponding TPSC value. This is done following a widely used method~\cite{Gukelberger_2015}
\begin{multline}
\Sigma^{\mathrm{TPSC}}(\vb{k}, ik_n) \rightarrow \Big[\Sigma^{\mathrm{TPSC}}(\vb{k}, ik_n) - \Sigma^{\mathrm{TPSC}}_{\mathrm{loc}}(ik_n)\Big]\\
 + \Sigma^{\mathrm{DMFT}}(ik_n).
 \label{eq:local_DMFT_self-energy_for_TPSC}
\end{multline}
The local TPSC self-energy is, by definition
\begin{equation}
\Sigma^{\mathrm{TPSC}}_{\mathrm{loc}}(ik_n) = \frac{1}{N}\sum_{\vb{k}}\Sigma^{\mathrm{TPSC}}(\vb{k}, ik_n),
\end{equation}
with $N$ the total number of $\vb{k}$-points in the wave vector grid for the TPSC calculation. The result should capture both the local effects of interactions from DMFT and the non-local effects from TPSC. The TPSC self-energy is obtained on a fine wavevector mesh. 

This approximation is easily combined with the previous modification to the method since the local DMFT self-energy and double occupancy are obtained from the same calculation. Both those steps make up the TPSC+DMFT method. It is important to note that this self-energy correction has no effect on the spin and charge susceptibilities, which are computed before the self-energy. In \emph{ansatz} TPSC (and TPSC+DMFT), the dressed Green's function $\Gfn^{(2)}$ is not used to compute new susceptibilities.
This is in contrast with the TPSC+ and TPSC+GG methods, where the process of calculating susceptibilities and self-energy is iterated until convergence \cite{tpscplus, Simard_2022, Schafer_2021}.

As a word of caution, we note that when the local TPSC self-energy is larger than the local DMFT self-energy, the subtraction in \eref{eq:local_DMFT_self-energy_for_TPSC} could lead to non-causal effects. We have not encountered this case, but we cannot exclude this possibility. Heuristically, the TPSC self-energy arises from long-wavelength collective modes, so it takes into account mostly non-local correlations rather than local ones. Its local component should generally be smaller than the local DMFT self-energy.   


\section{Results}
\label{section:results}
Here we compare benchmarks with TPSC and TPSC+DMFT.
We start in ~\secref{section:results_double_occ} with the double occupancy that, in the first step of TPSC, determines the spin and charge susceptibilities. 
We then compare in turn our results with DiagMC benchmarks for the spin susceptibility in ~\secref{section:results_spin_susceptibility}, the corresponding magnetic correlation length in ~\secref{section:results_magnetic_corr_length} and the charge susceptibility in~\secref{section:results_charge_susceptibility}. 
We end with the self-energy in~\secref{section:results_self_energy}.

\subsection{Double occupancy}
\label{section:results_double_occ}

\begin{figure}[htb]
\centering
\includegraphics[width=\figwidth]{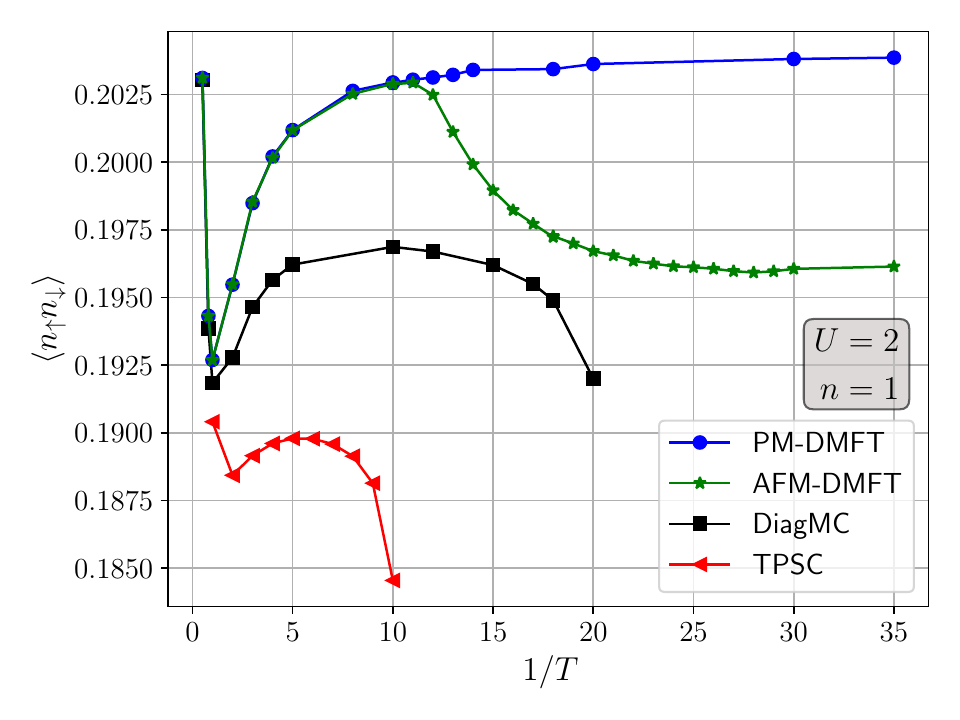}
\caption{Double occupancy as a function of inverse temperature at half filling for TPSC, for the reference exact DiagMC~\cite{Rossi_2017} results  Ref.~\cite{Schafer_2021} and for the paramagnetic and antiferromagnetic versions of DMFT. Error bars on DMFT results are smaller than the markers. The DiagMC results fall generally between TPSC and DMFT. Mathematically, the decrease in TPSC double occupancy at half filling is a consequence of the fact that the susceptibility $\chi^{(1)}(T)$ diverges as $T$ decreases so that $U_{\mathrm{sp}}$ in the interacting susceptibility \eref{eq:chi_sp} must vanish. The \emph{ansatz} \eref{eq:ansatz_tpsc} then automatically implies that double occupancy must vanish at $T=0$.}
\label{fig:comp_D_beta_demi_rempli}
\end{figure}

 The double occupancy at half filling and small interaction strength $U=2$ for both TPSC and DMFT is shown in \figref{fig:comp_D_beta_demi_rempli}, as a function of inverse temperature. 
Both PM-DMFT and AFM-DMFT overestimate the double occupancy when compared with the DiagMC results for $T<1$, while TPSC underestimates the double occupancy over the whole range of temperatures. 
For all methods, the discrepancies over the available range of data with DiagMC are of order $5\%$, relatively small, but they can have important consequences on susceptibilities. 

At temperatures lower than $T\approx \flatfrac{1}{12}$, the double occupancy calculated from AFM-DMFT deviates from PM-DMFT because this is the DMFT Néel temperature, below which AFM-DMFT predicts long-range antiferromagnetic order. 
This ordering leads to a more accurate double occupancy, reproducing a similar downturn at $T < T_{\mathrm{DMFT}}^N$ to DiagMC.
Long-range ordering at finite temperature is forbidden in two-dimensional systems by the Mermin-Wagner theorem, but long-wavelength spin fluctuations play a similar role in lowering the double occupancy at low temperatures.
This type of long-range fluctuations is not possible by design in single-site DMFT, which is why the PM-DMFT double occupancy saturates at low temperature without downturn.

\begin{figure}[htb]
\centering
\includegraphics[width=\figwidth]{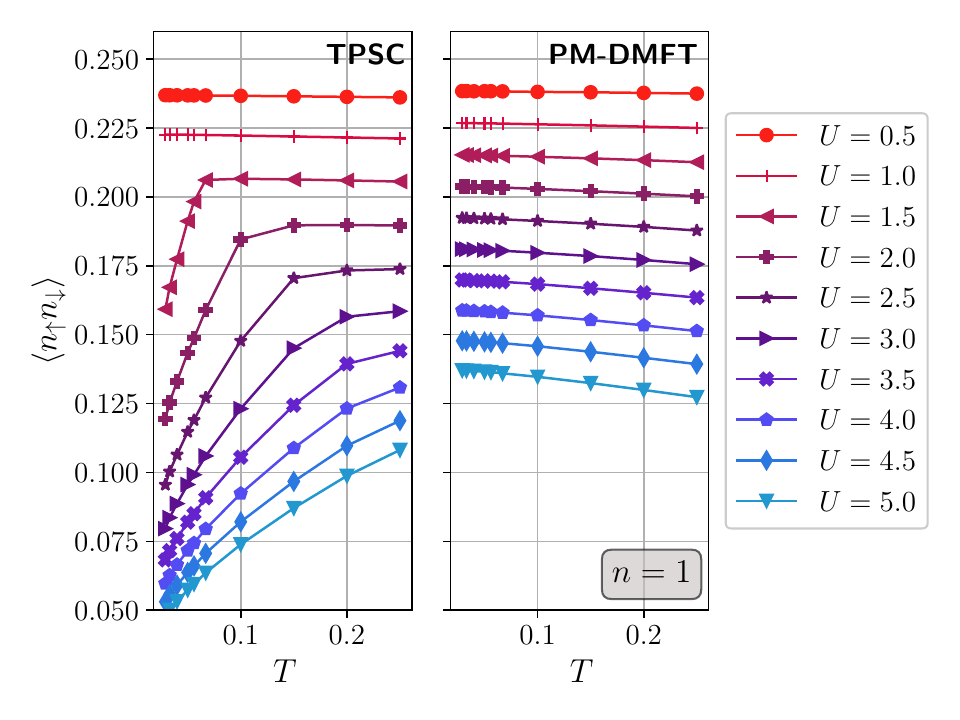}
\caption{Double occupancy for TPSC and for PM-DMFT at half filling, as a function of the temperature of the model for different values of interaction $U$. The low temperature fall signals the regime where the antiferromagnetic spin correlation length increases, leading to a characteristic spin fluctuation frequency smaller than temperature, the so-called renormalized classical regime.
Uncertainty in the CT-HYB evaluation of double occupancy in DMFT leads to error bars on PM-DMFT results smaller than the markers.}
\label{fig:comp_D_T_U_simkovic_n_1.000}
\end{figure}

A more complete comparison between the TPSC and TPSC+DMFT double occupancy at half filling is shown in \figref{fig:comp_D_T_U_simkovic_n_1.000}. 
In the case of TPSC+DMFT, the double occupancy is in fact that of PM-DMFT.
It can be seen clearly in the figure that, for $U \gtrsim 1.5$, the TPSC double occupancy sharply decreases at low enough temperature, whereas the PM-DMFT double occupancy saturates to a finite value. 
Since spin correlations increase at low temperature, the local moment  $\expval{(S^z)^2}$ must increase so that, from the local moment sum rule~\eref{eq:sumrule_sp}, the double occupancy must decrease. 
The decrease signals the beginning of the renormalized classical regime. 
This decrease is also seen in the DiagMC results in \figref{fig:comp_D_beta_demi_rempli}, but at lower temperature than in TPSC. 
By contrast, DMFT is oblivious to this physics, so the double occupancy saturates. 
The underestimation of the double occupancy by TPSC and overestimation by DMFT can lead to compensations that are to our advantage as we will see later.  
As will be discussed in \secref{section:results_spin_susceptibility} on the spin susceptibility, the feedback between double occupancy and spin fluctuations occurs differently in TPSC and TPSC+DMFT.

\begin{figure}[htb]
\centering
\includegraphics[width=\figwidth]{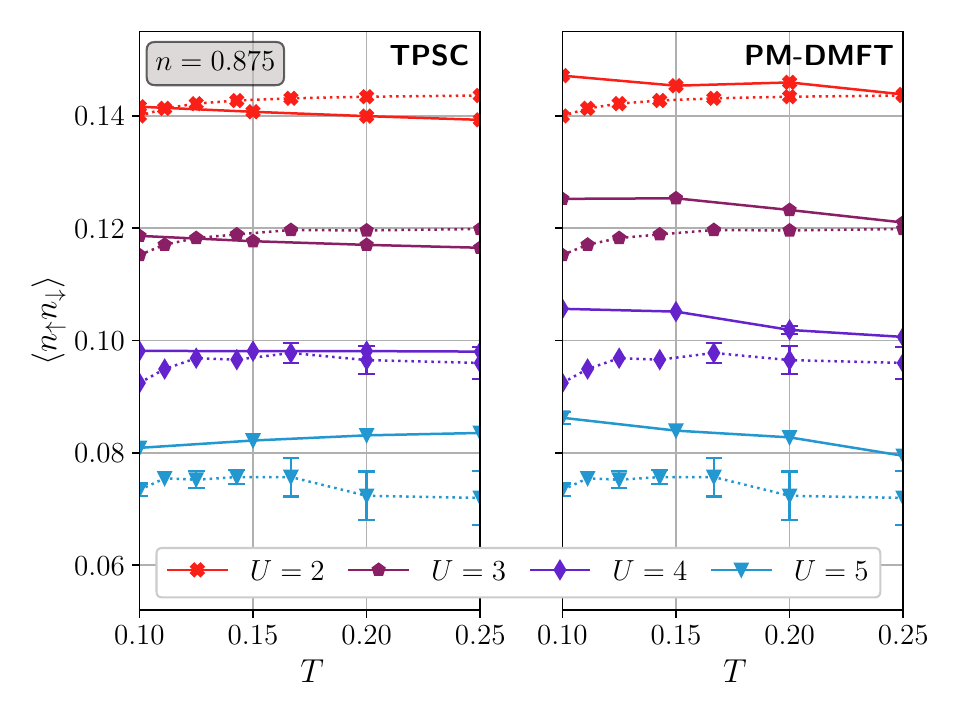}
\caption{Double occupancy at $n=0.875$, as a function of the temperature for different values of the interaction $U$. 
The benchmark DiagMC results are reproduced from \cite{Simkovic_2022_benchmark} with error bars and dotted lines. 
TPSC generally underestimates the benchmark while PM-DMFT overestimates the benchmark. 
TPSC agrees better with the benchmark for $U<5$.
Uncertainty in the CT-HYB evaluation of the density and double occupancy in DMFT leads to error bars on PM-DMFT results, smaller than 0.002. 
A few are shown, for example at $U= 4$, $T=0.2$.}
\label{fig:comp_D_T_U_simkovic_n_0.875}
\end{figure}

Figure \ref{fig:comp_D_T_U_simkovic_n_0.875} shows the comparisons away from half filling at $n=0.875$. 
TPSC still slightly underestimates the double occupancy while PM-DMFT overestimates it. In general, TPSC  gives a slightly better agreement with the benchmark DiagMC results up to $U=4$.
This regime is sufficiently far from the renormalized classical regime that the TPSC and benchmark double occupancies do not decrease significantly at low temperature. 

\begin{figure}[htb]
\centering
\includegraphics[width=\figwidth]{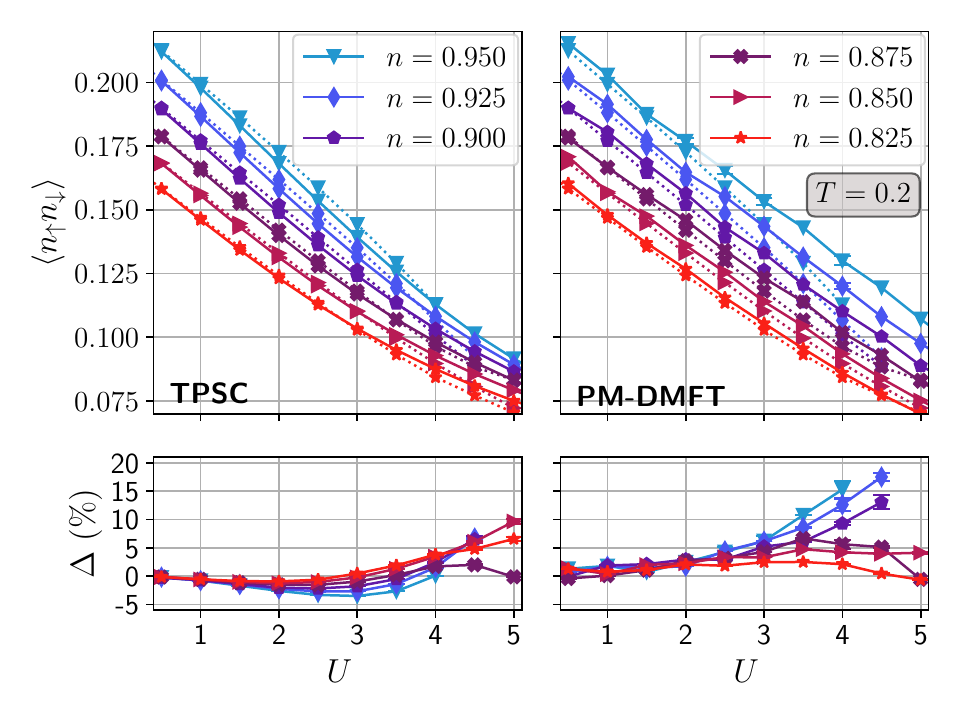}
\caption{Top row: Double occupancy at $T=0.2$ as a function of the interaction strength for various fillings. Benchmark DiagMC results are reproduced from Ref.~\cite{Simkovic_2022_benchmark} with error bars and dotted lines. 
TPSC generally underestimates the benchmark while PM-DMFT overestimates the benchmark. 
TPSC agrees better with the benchmark for $U<5$.
Uncertainty in the CT-HYB evaluation of the density and double occupancy in PM-DMFT leads to typical errors between $1\%$ and $5\%$. 
A few are indicated, for example at $n=0.95$, $U=4$. Bottom row: Relative deviation of the results from the exact DiagMC values.}
\label{fig:comp_D_U_n_simkovic}
\end{figure}

Again far from the renormalized classical regime at temperature $T=0.2$, the dependence on $U$ of the double occupancy for various densities is shown in \figref{fig:comp_D_U_n_simkovic}. 
Here again, one notices the general trend that TPSC underestimates and PM-DMFT overestimates the benchmark, with TPSC agreeing overall better with the benchmarks.

\subsection{Spin susceptibility}
\label{section:results_spin_susceptibility}

\begin{figure*}[hbt]
\centering
\includegraphics[width=\linewidth]{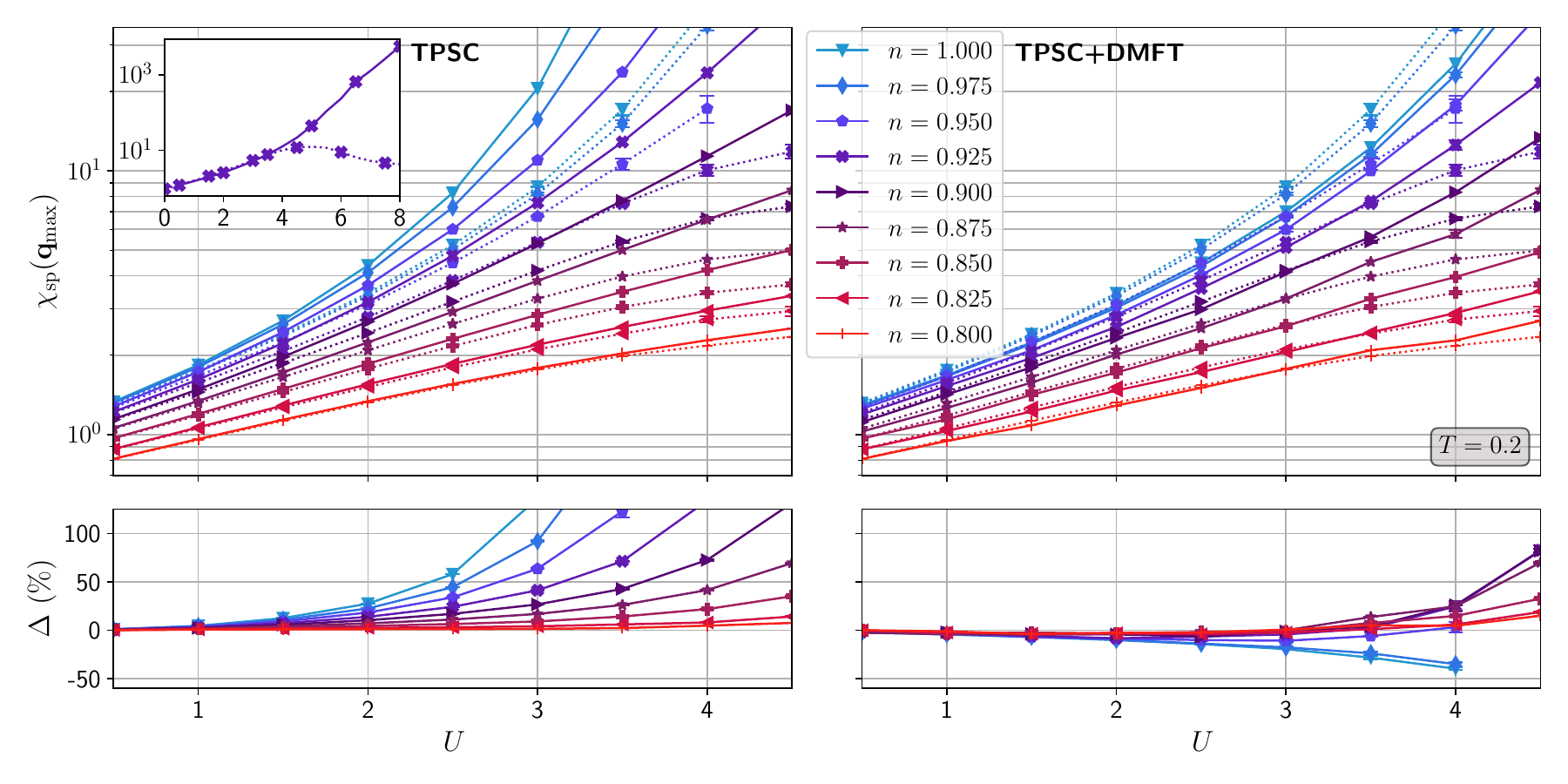}
\caption{Top row: Semilogarithmic plot of the dependence on interaction strength of the maximum of the spin susceptibility at $T=0.2$ for various densities for both TPSC and TPSC+DMFT.
Benchmark DiagMC results from Ref.~\cite{Simkovic_2022_benchmark} are displayed with error bars and dotted lines. 
TPSC+DMFT generally agrees better with DiagMC because TPSC enters the renormalized classical regime at too high temperature.
Uncertainty in the CT-HYB evaluation of the density and double occupancy in DMFT leads to typical errors from $1\%$ away from half filling and low $U$ up to $6\%$ at $n=0.975$ and $U=5$. 
Inset: Typical DiagMC and TPSC results extended to $U=8$ to show that both TPSC methods miss Heisenberg-Mott physics. Bottom row: Relative deviation of the results from the exact DiagMC values.}
\label{fig:comp_chispmax_U_n}
\end{figure*}

From the double occupancy, we can now evaluate the spin fluctuations. As a proxy for the accuracy of the spin susceptibility at vanishing Matsubara frequency,  \figref{fig:comp_chispmax_U_n} displays on a semilogarithmic plot the maximum of the spin susceptibility at temperature $T=0.2$. 
In the small interaction strength regime $0<U<2$, the results agree quite well with the benchmark DiagMC results \cite{Simkovic_2022_benchmark} for both methods~\footnote{A different definition of the spin operator was used in \cite{Simkovic_2022_benchmark}.The Monte Carlo spin susceptibility was thus scaled by a factor of 4 to compare with the results from both TPSC methods.}. 
In the intermediate regime $2<U<4$, TPSC+DMFT gives quantitatively better results, especially close to half filling. 
There, standard TPSC is at its worse because, as seen in  \figref{fig:comp_D_beta_demi_rempli}, it enters the renormalized classical regime at too high temperature. 
This also affects the charge fluctuations, as we discuss later.

\begin{figure}[htb]
\centering
\includegraphics[width=\figwidth]{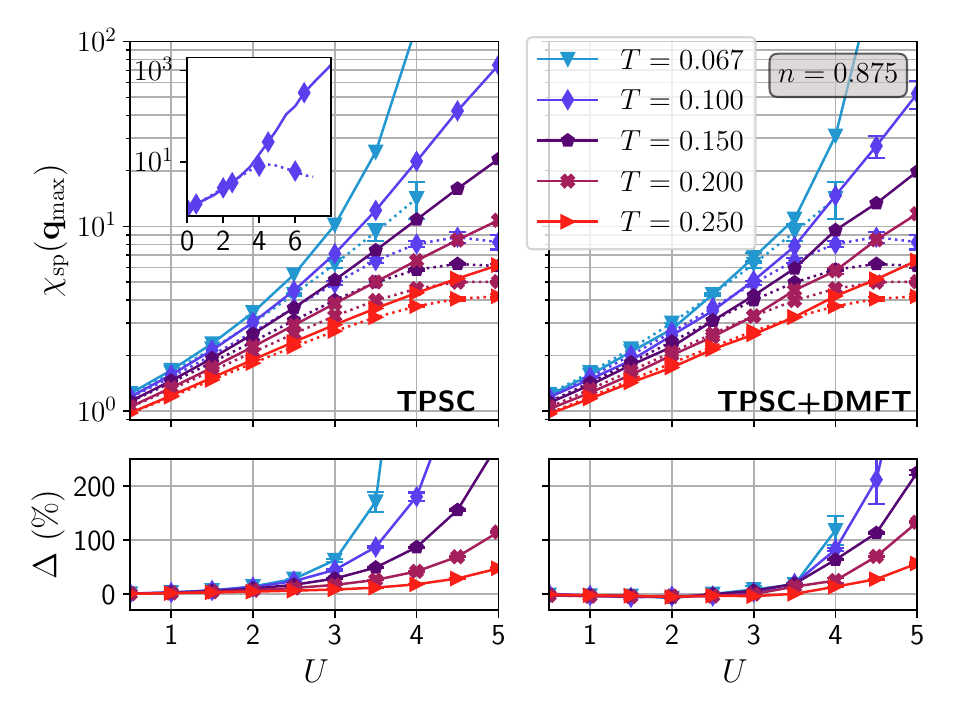}
\caption{Top row: Semilogarithmic plot of the dependence on interaction strength of the maximum of the spin susceptibility at filling $n=0.875$ for various temperatures for both TPSC and TPSC+DMFT. 
Benchmark DiagMC results from Ref.~\cite{Simkovic_2022_benchmark} are displayed with error bars and dotted lines. TPSC+DMFT generally agrees better with DiagMC because TPSC enters the renormalized classical regime at too high temperature.
At low temperature, even TPSC+DMFT starts to deviate substantially from benchmark results for $U$ larger than $3$. 
Uncertainty in the CT-HYB evaluation of the density and double occupancy in DMFT leads to typical errors from $2\%$-$4\%$ at high temperature and low $U$ up to $20\%$ at $T=0.1$ and $U=5$. 
A few are indicated.  
Inset: Typical DiagMC and TPSC results extended to $U=8$ to show that both TPSC methods miss Heisenberg-Mott physics. Bottom row: Relative deviation of the results from the exact DiagMC values.}
\label{fig:comp_chispmax_U_T}
\end{figure}

A similar behavior at the fixed density $n=0.875$ for various temperatures is seen in the semilogarithmic plot of the maximum spin susceptibility as a function of interaction $U$ in \figref{fig:comp_chispmax_U_T}. 
The low-interaction maximum spin susceptibility agrees with benchmark DiagMC results for both methods, with TPSC+DMFT in slightly better quantitative agreement with DiagMC for $2<U<4$. 
This is especially true for low temperatures, where TPSC overestimates the spin susceptibility a lot more than TPSC+DMFT, again because it enters the renormalized classical regime at too high temperature. 
The system being relatively far from half filling, the improvement brought by TPSC+DMFT is less than that at $n$ close to half filling.
This is consistent with the fact that the TPSC \emph{ansatz}, being a generalized Hartree-Fock approximation, is better away from half filling in the dilute limit. 

In the inset of both \figref{fig:comp_chispmax_U_T} and \figref{fig:comp_chispmax_U_n}, we show that the $U>5$ regime is not well approximated by either TPSC method.
The DiagMC spin susceptibility starts to decrease with increasing $U$ since the system enters the Heisenberg regime where local moments and Mott physics are prevalent.
In this regime, the large interaction strength leads to a saturation of the local magnetic moments $\ev{S_z^2}$, and the Hubbard model is well approximated by a $t-J$ model, with a super-exchange parameter $J=4t^2/U$.
In this regime, increasing $U$ decreases super-exchange and suppresses spin fluctuations.
It is also in this strong-interaction regime that a Mott gap opens, which is why we refer to this phenomenon as Heisenberg-Mott physics.
This is completely missed by both TPSC methods, which predict an ever-increasing spin susceptibility with interaction strength.

\begin{figure}[htb]
\centering
\includegraphics[width=\figwidth]{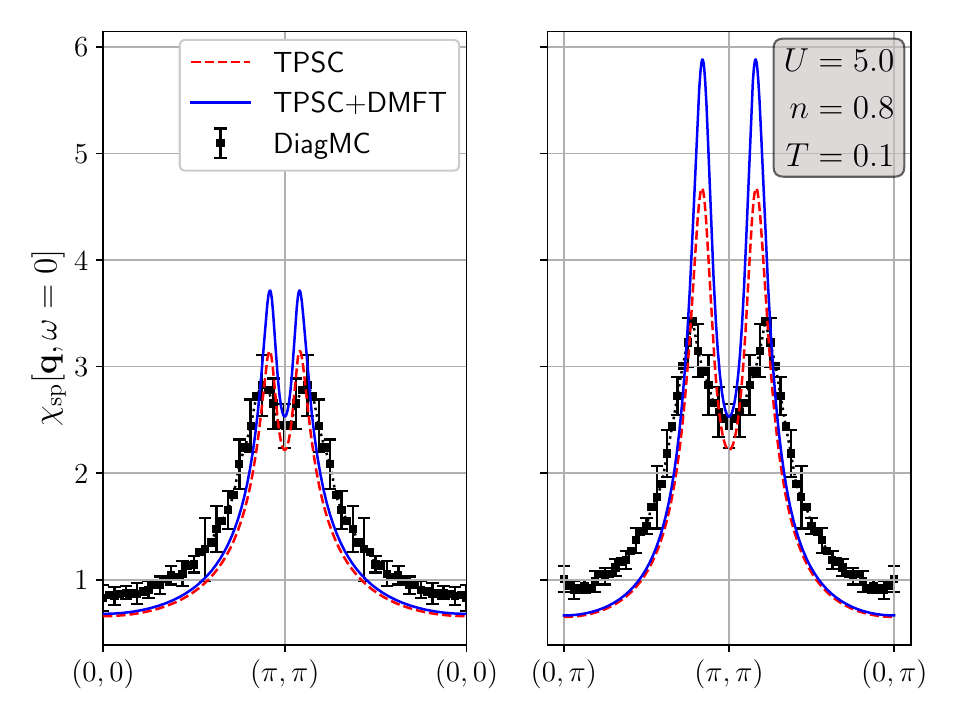}
\caption{Wave-vector dependence of the spin susceptibility at $U=5$, $T=0.1$, and $n=0.8$ along the zone diagonal (left panel) and boundary (right panel). Monte-Carlo results are reproduced from \cite{Simkovic_2022_benchmark} with error bars and dotted lines. For these parameters, we are beyond the limit of validity of both TPSC methods. The spin fluctuations in TPSC are larger than the benchmark because in weak interaction methods, these fluctuations increase indefinitely with $U$ instead of entering the regime where superexchange becomes important.}
\label{fig:comp_chisp_Q_path}
\end{figure}

The wave-vector dependence of the spin susceptibility is pictured along the Brillouin zone diagonal and edge in \figref{fig:comp_chisp_Q_path} for $n=0.8$  when $U$ is $5$ and $T$ is $0.1$. This is at the beginning of the renormalized classical regime where the spin fluctuations start to grow for all methods. 
The inset of \figref{fig:comp_chispmax_U_n} shows that for $U=5$ the system is in a regime where we expect  both TPSC methods to start breaking down.
This is partly the case, with TPSC and TPSC+DMFT overestimating the spin susceptibility. 
TPSC is closer to the DiagMC results for this set of parameters. 
Nevertheless, the qualitative behavior is correct because $U=5$ is at the border of where both TPSC methods fail. 
Note that in the long correlation length limit, the double incommensurate peaks should appear only along the zone edge, not along the diagonal.
The weight along the diagonal occurs because the correlation length is still small so that it influences a large almost circular region around $(\pi,\pi)$, as can clearly be seen from the color plot on the left of Fig.~2 of Ref.~\cite{Simkovic_2022_benchmark}.  

\begin{figure}[htb]
\centering
\includegraphics[width=\figwidth]{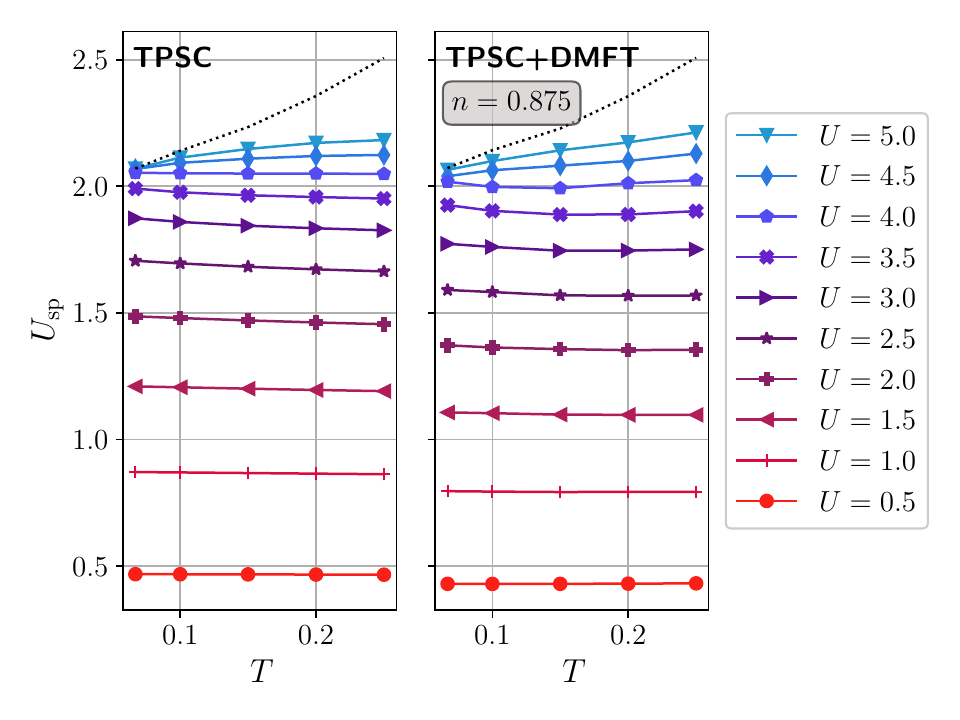}
\caption{Spin vertex obtained by both TPSC methods at $n=0.875$, as a function of the bare interaction strength $U$ and temperature. The dotted black line represents the critical value $U_{\mathrm{crit}}(T)$ of the spin vertex, for which $\chi_{\mathrm{sp}}$ diverges.
Error bars on TPSC+DMFT data coming from the CT-HYB evaluation of the density in DMFT are smaller than the markers and not shown.}
\label{fig:comp_Usp_T_U_n_0.875}
\end{figure}

The difference in the spin susceptibilities of the two TPSC methods originates from the different renormalized irreducible spin vertices  $U_{\mathrm{sp}}$. 
It is thus instructive to plot $U_{\mathrm{sp}}$ for the two methods. 
Away from half filling, at $n=0.875$, we see in \figref{fig:comp_Usp_T_U_n_0.875} that the value of the spin vertex is somewhat smaller in TPSC+DMFT than in TPSC. 
This is consistent with the slightly larger double occupancy from DMFT seen in \figref{fig:comp_D_T_U_simkovic_n_0.875} for most values of $U$. 

\begin{figure}[htb]
\centering
\includegraphics[width=\figwidth]{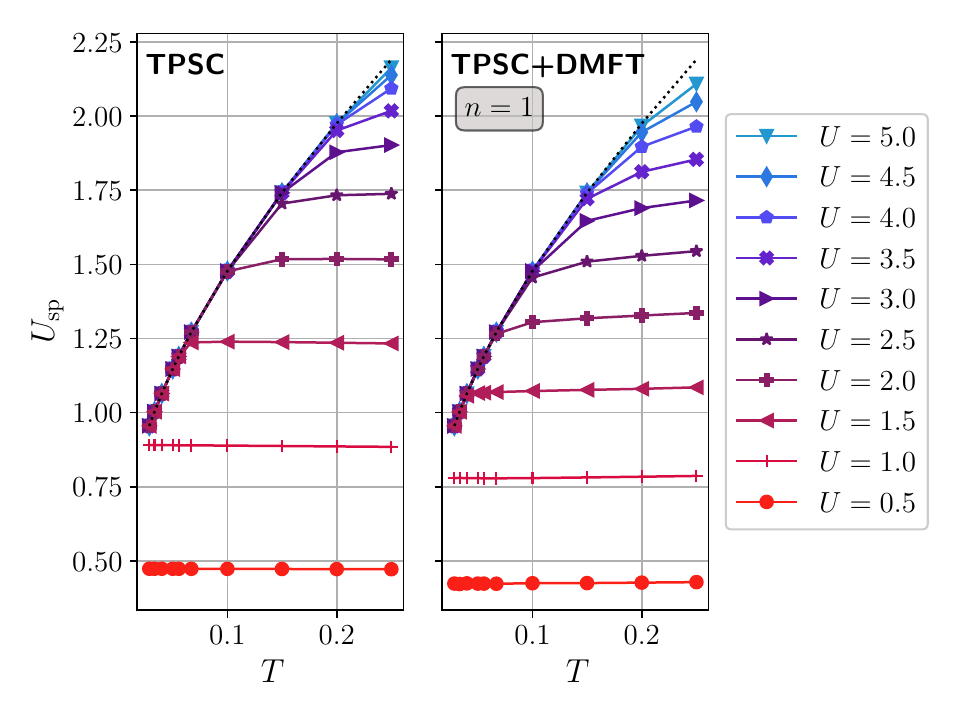}
\caption{Spin vertex obtained from TPSC and TPSC+DMFT at half filling ($n=1$), as a function of the bare interaction strength $U$ and temperature. The dotted black line represents the critical value $U_{\mathrm{crit}}(T)$ of the spin vertex, for which $\chi_{\mathrm{sp}}$ diverges. At half filling, the uncertainty on the density is negligible, which leads to no significant error bars on TPSC+DMFT results.}
\label{fig:comp_Usp_T_U_n_1.000}
\end{figure}

In general, to satisfy the Mermin-Wagner theorem, $U_{\mathrm{sp}}\chi^{(1)}(q)/2$ must not equal unity. We thus define a critical $U_{\mathrm{sp}}$ by $U_{\mathrm{crit}}(T)=2/\chi^{(1)}_{\mathrm{max}}(q)$.
At half filling, the divergence of $\chi^{(1)}(q)$ at $\vb{Q}=(\pi, \pi)$ forces $U_{\mathrm{sp}}$ to vanish at $T\rightarrow 0$. This is where the difference between the two methods is more pronounced. 
Since $\chi^{(1)}$ diverges in the same way for low temperature in both methods, the critical value of $U_{\mathrm{sp}}$ is identical in both methods. In \figref{fig:comp_Usp_T_U_n_0.875} and \figref{fig:comp_Usp_T_U_n_1.000}, $U_{\mathrm{crit}}(T)$ corresponds to the dotted black line. 
The temperature at which the spin vertex starts to decrease can be identified with the temperature where $U_{\mathrm{crit}}(T)$ is approximately equal to the high-temperature value $U_{\mathrm{sp}}^{\mathrm{HT}}$ of $U_{\mathrm{sp}}$.
As seen in \figref{fig:comp_Usp_T_U_n_1.000}, for a given $U$, the spin vertex starts to decrease with $T$ at a higher temperature for TPSC than for TPSC+DMFT, consistent with the lower value of $U_{\mathrm{sp}}^{\mathrm{HT}}$ in TPSC+DMFT. 

\subsection{Magnetic correlation length}
\label{section:results_magnetic_corr_length}

\begin{figure}[htb]
\centering
\includegraphics[width=\figwidth]{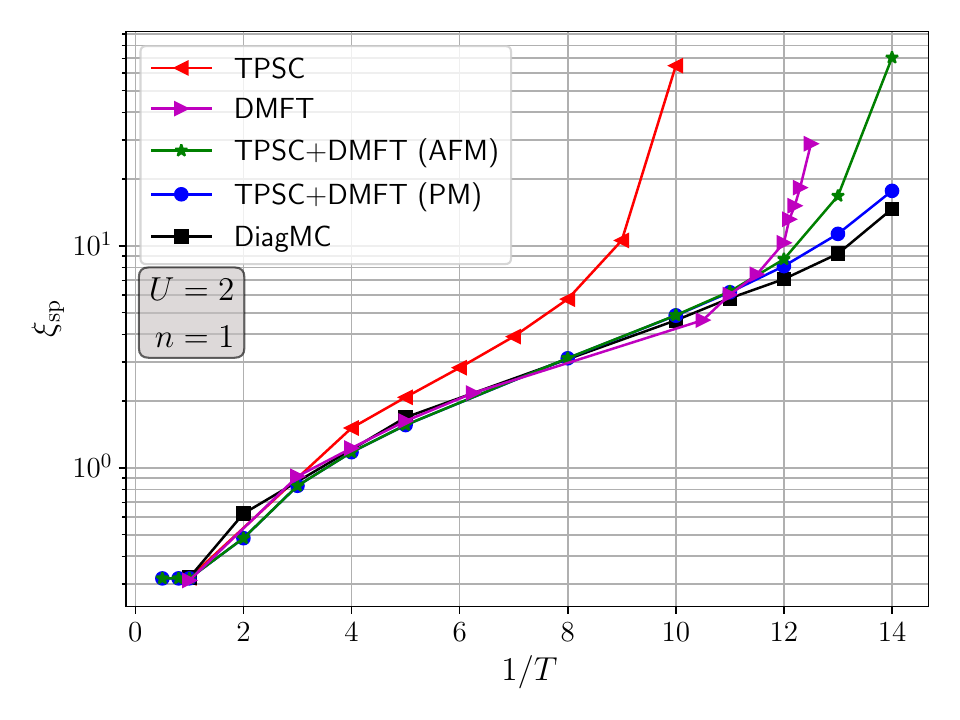}
\caption{Magnetic correlation length of the model at half filling for various techniques, as a function of inverse temperature. DiagMC and DMFT results are reproduced from Ref. \cite{Schafer_2021}. TPSC+DMFT in the paramagnetic state gives the best agreement with DiagMC.}
\label{fig:comp_xisp_beta}
\end{figure}

The magnetic correlation length obtained from a fit, Eq.~\ref{eq:fit_lorentzienne_xisp},  of the spin susceptibility at half filling is shown in \figref{fig:comp_xisp_beta}. For the TPSC+DMFT results, both paramagnetic and antiferromagnetic DMFT were tested. While the TPSC correlation length starts to deviate from the Monte-Carlo result at the relatively high temperature of $T=\flatfrac{1}{5}$, the values obtained by both TPSC+DMFT methods closely match the exact results above $T\approx\flatfrac{1}{12}$. The double occupancy from paramagnetic DMFT gives a better magnetic correlation length than the AFM-DMFT double occupancy, tracking the DiagMC value qualitatively to the lowest temperature available.

The magnetic correlation length extracted directly from DMFT (calculated in \cite{Schafer_2021}) is also shown on that figure. It is closer to the exact result than TPSC, but farther than both TPSC+DMFT methods. It therefore seems that the combination of both local effects from the DMFT double occupancy and non-local effects from TPSC gives the best representation of the physics in that case. 

\subsection{Charge susceptibility}
\label{section:results_charge_susceptibility}

In both TPSC methods, the Pauli principle controls the relative values of spin and charge susceptibilities. In other words, the sum of the local spin sum rule \eref{eq:sumrule_sp} and of the local charge sum rule \eref{eq:sumrule_ch} is independent of interaction strength $U$. It depends only on the filling $n$ because of the Pauli principle. Hence, when interactions increase spin fluctuations, charge fluctuations must decrease. This is important for understanding how the results of this section are related to results on the spin susceptibility.

\begin{figure}[htb]
\centering
\includegraphics[width=\figwidth]{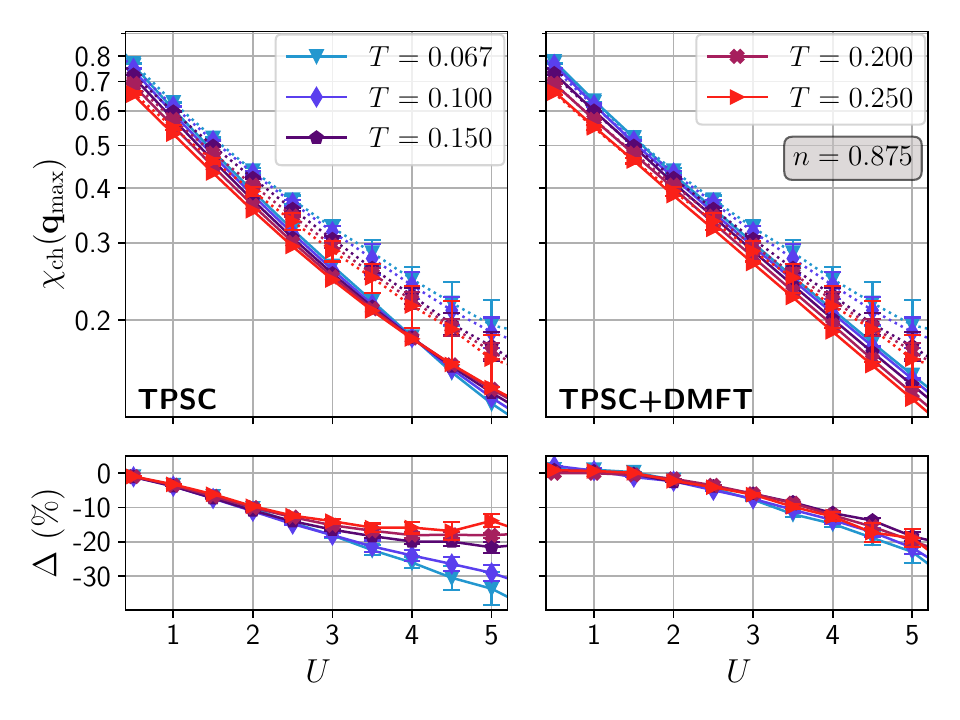}
\caption{Top row: Maximum of the charge susceptibility obtained from TPSC and TPSC+DMFT, as a function of the bare interaction strength $U$ for various temperatures, at $n=0.875$ (solid lines). Corresponding benchmark DiagMC results from Ref.~\cite{Simkovic_2022_benchmark} are displayed with error bars and dotted lines. 
Uncertainty on the evaluation of the CT-HYB density and double occupancy leads to error bars on TPSC+DMFT results smaller than $0.7\%$ and of the size of the markers. Bottom row: Relative deviation of the results from the exact DiagMC values.}
\label{fig:comp_chichmax_U_T}
\end{figure}

As a proxy for the accuracy of the charge susceptibility at vanishing Matsubara frequency, \figref{fig:comp_chichmax_U_T}  displays on a semilogarithmic plot the maximum of the charge susceptibility for various temperatures as a function of interaction strength at filling $n=0.875$. 
In the small interaction strength regime $0 < U < 2$ the results agree quite well with the benchmark DiagMC results \cite{Simkovic_2022_benchmark} for both methods, with TPSC+DMFT showing slightly better agreement. 
In the intermediate regime $2 < U < 4$, TPSC+DMFT gives quantitatively better results. 
Both TPSC methods underestimates the charge susceptibility, as expected from the Pauli principle and the overestimation of the spin fluctuations in \figref{fig:comp_chichmax_U_T}. 
The temperature dependence of $\chi_{\mathrm{ch}}(\vb{q}_{\mathrm{max}})$ for $U\gtrsim4$ has a behavior opposite to that of the DiagMC result, while the TPSC+DMFT temperature dependence stays consistent.

\begin{figure}[htb]
\centering
\includegraphics[width=\figwidth]{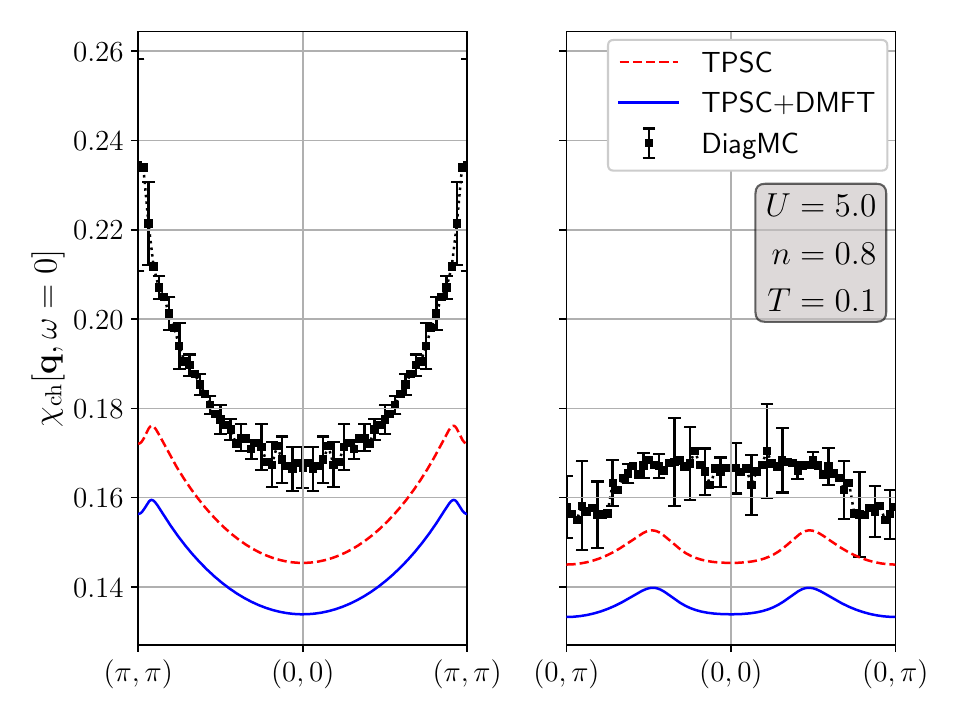}
\caption{Wave-vector dependence of the charge susceptibility at $U=5$, $T=0.1$, and $n=0.8$. 
DiagMC results are reproduced from \cite{Simkovic_2022_benchmark}.
The better agreement of TPSC with DiagMC is consistent through the Pauli principle with the better agreement found for the spin fluctuations in \ref{fig:comp_chisp_Q_path}.}
\label{fig:comp_chich_Q_path}
\end{figure}

The wave-vector dependence of the charge susceptibility along the zone diagonal and zone edge is shown in \figref{fig:comp_chich_Q_path} for $n=0.8$, $U=5$ and $T=0.1$. As mentioned for spin fluctuations, the system is in a regime where we expect TPSC to start breaking down. 
Since both TPSC methods overestimate spin fluctuations, as seen in ~\figref{fig:comp_chisp_Q_path}, the Pauli principle forces the charge fluctuations to be underestimated compared to DiagMC, as was also the case in \figref{fig:comp_chichmax_U_T}. Similarly, the better agreement of TPSC with DiagMC spin fluctuations leads to better agreement of TPSC with DiagMC charge fluctuations through the Pauli principle.

\begin{figure}[htb]
\centering
\includegraphics[width=\figwidth]{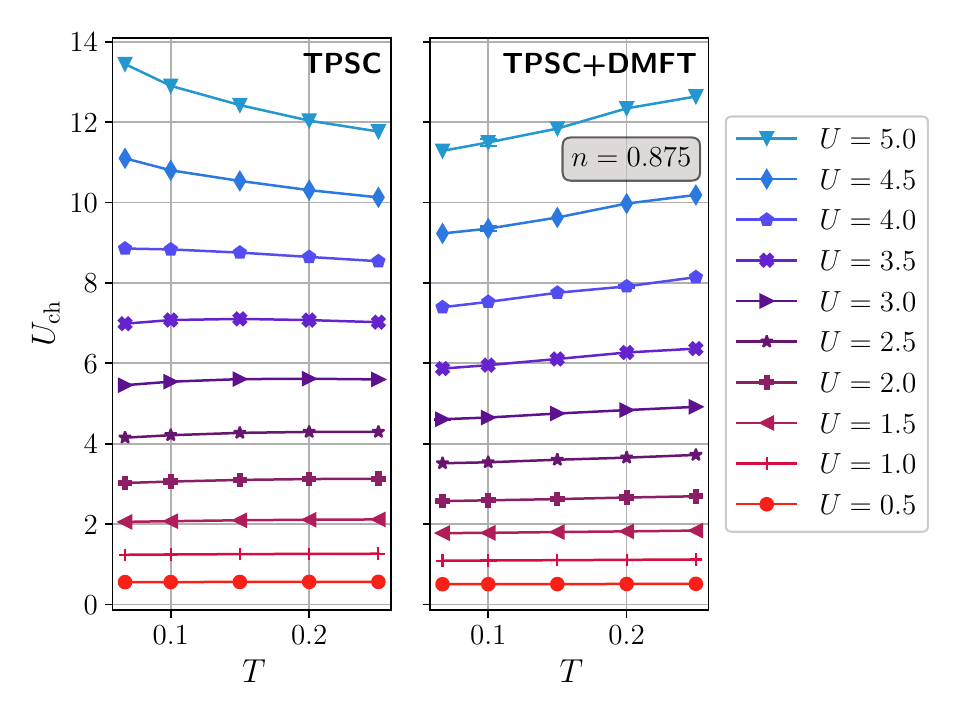}
\caption{Charge vertex obtained by TPSC and TPSC+DMFT at $n=0.875$, as a function of the bare interaction strength $U$ and temperature.
Error bars caused by the uncertainty on the density are smaller than the markers.}
\label{fig:comp_Uch_T_U_n_0.875}
\end{figure}

\begin{figure}[htb]
\centering
\includegraphics[width=\figwidth]{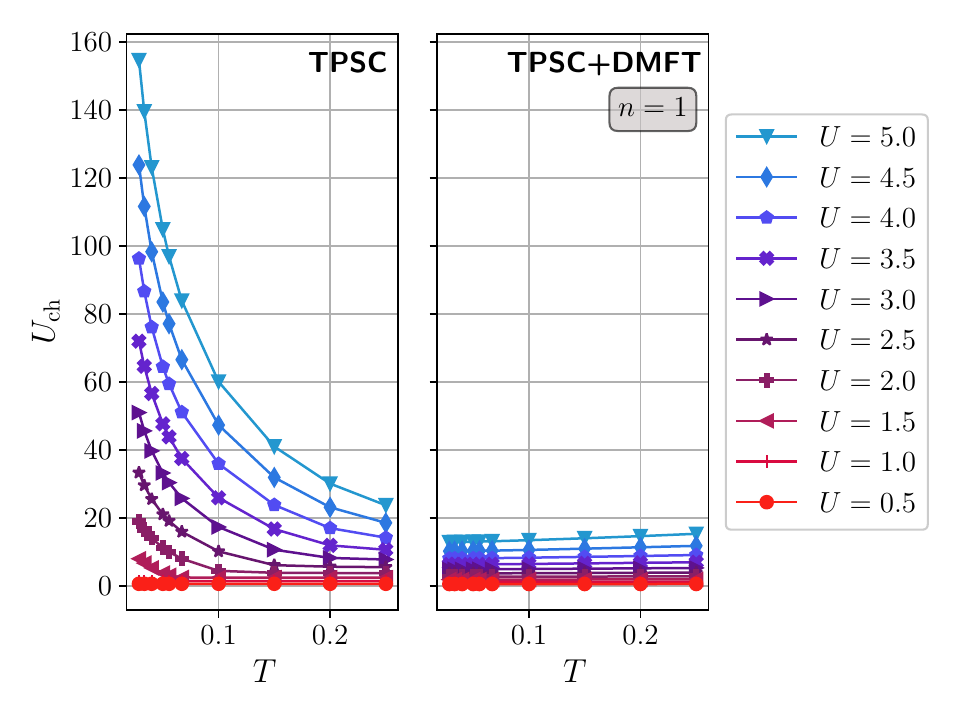}
\caption{Charge vertex obtained by TPSC and TPSC+DMFT at half filling, as a function of the bare interaction strength $U$ and temperature.
At low temperature, the downturn of the double occupancy in TPSC coming from the \emph{ansatz} suppresses the charges fluctuations, in contrast with TPSC+DMFT, where they stay similar no matter the temperature.}
\label{fig:comp_Uch_T_U_n_1.000}
\end{figure}

The difference in the charge susceptibilities of the two TPSC methods originates from the different renormalized irreducible charge vertices $U_{\mathrm{ch}}$.
We thus plot $U_{\mathrm{ch}}$ as a function of temperature for the two methods in \figref{fig:comp_Uch_T_U_n_0.875} and \figref{fig:comp_Uch_T_U_n_1.000} for $n=0.875$ and $n=1$ respectively. 
Away from half filling and for $U\gtrsim 4$, the temperature dependence of $U_{\mathrm{ch}}$ computed in TPSC deviates from the TPSC+DMFT charge vertex. 
This reflects the change in the temperature dependence of the double occupancy shown in \figref{fig:comp_D_T_U_simkovic_n_0.875}.
The half-filling case shown in \figref{fig:comp_Uch_T_U_n_1.000} is an even more extreme example of the difference in charge vertex calculated by both methods. 
This occurs because in TPSC the \emph{ansatz} forces the double occupancy to vanish at low temperature with $U_{\mathrm{sp}}$. 
Since the right-hand side of the local charge sum rule in \eref{eq:sumrule_ch} is proportional to double occupancy when $n=1$, the charge fluctuations must be completely suppressed by $U_{\mathrm{ch}}$ in this situation.

\subsection{Self-energy}
\label{section:results_self_energy}

\begin{figure*}[htb]
\centering
\includegraphics[width=0.9\linewidth]{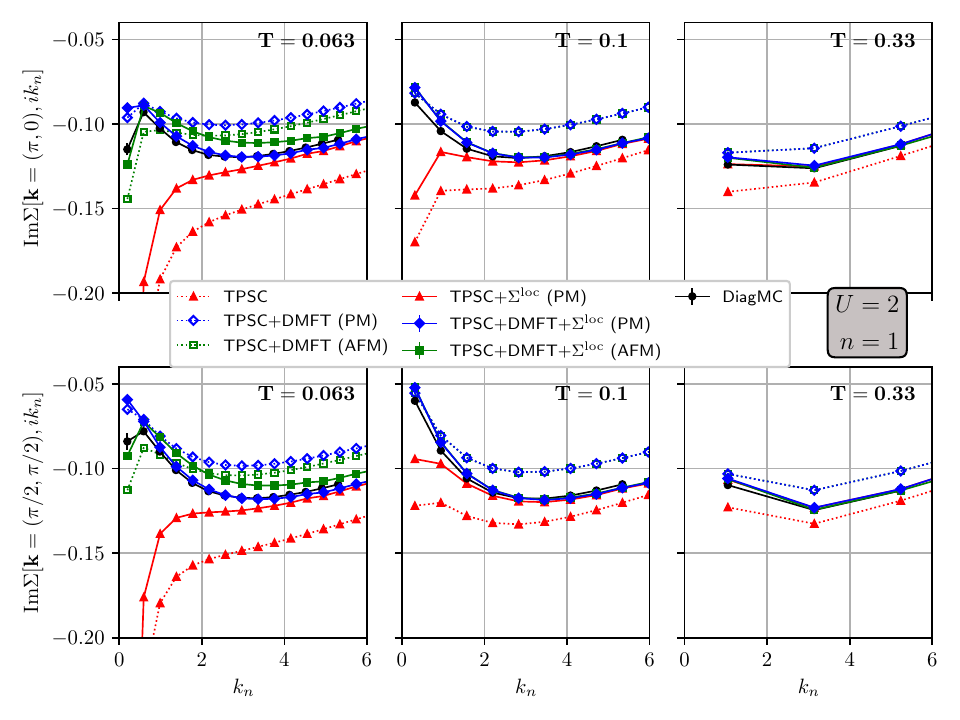}
\caption{Imaginary part of the self-energy as a function of Matsubara frequency $k_n$ at the antinode $\vb{k}=(\pi,0)$ in the top row and at the node $\vb{k}=(\frac{\pi}{2},\frac{\pi}{2})$ in the bottom row. 
The lowest temperatures are to the left. 
All calculations were done at interaction strength $U=2$ and half filling $n=1$. 
Error bars on DiagMC and $\Sigma^{\mathrm{loc}}$ methods are displayed but usually smaller than the markers.
For clarity, the curves for which the substitution of the local part of the self-energy was done are explicitly indicated (in solid lines).
There is a difference between PM-DMFT and AFM-DMFT methods only below the Néel temperature, namely in the leftmost panels.
The combination of paramagnetic TPSC+DMFT with the corresponding $\Sigma^{\mathrm{loc}}$ gives the best overall agreement with 
DiagMC data taken from \cite{Schafer_2021}.}
\label{fig:comp_self_energy_demi_rempli_U_2}
\end{figure*}

In both TPSC methods, the self-energy is computed after the spin and charge susceptibilities with Eq.~\ref{eq:self-energie_tpsc}. 
In Matsubara frequencies, it is the imaginary part of the self-energy that is most revealing of the Fermi liquid vs non-Fermi liquid properties. 
It is shown for $U=2$ and three temperatures at half filling in \figref{fig:comp_self_energy_demi_rempli_U_2}. 
The  antinodal $\vb{k}=(\pi,0)$ point is in the top row and the nodal $\vb{k}=(\tfrac{\pi}{2},\tfrac{\pi}{2})$ point in the bottom row.
In that figure, the red dotted line is for TPSC, while TPSC+DMFT methods are shown in dotted blue lines when double occupancy is obtained from PM-DMFT and in dotted green lines when double occupancy is obtained from AFM-DMFT.
The corresponding solid lines correspond to the $\Sigma^{\mathrm{loc}}$ approximation of \secref{section:methods} where the local part of the TPSC self-energy is replaced by the local part of the DMFT self-energy.
The DiagMC results from Ref.~\cite{Schafer_2021} are in black. 

For $T=0.063$ and $T=0.1$, on the left and middle panels respectively, the TPSC+DMFT self-energies are much closer to the DiagMC results than TPSC. 
Taking the positive slope of the self-energy between the first two Matsubara frequencies as a proxy to the beginning of the pseudogap, it is apparent that TPSC enters this regime at a higher temperature than DiagMC and TPSC+DMFT.

The $\Sigma^{\mathrm{loc}}$ approximation in conjunction with TPSC+DMFT gives the best self-energies (solid green and blue lines) especially at low temperature, as we now show. 
The high-frequency behavior of the self-energy is almost identical to the DiagMC self-energy if we use PM-DMFT. 
Below the DMFT Néel temperature, the paramagnetic or antiferromagnetic solutions both give results similar to DiagMC, but the entry in the pseudogap regime of AFM-DMFT is in better agreement with DiagMC for both nodal and antinodal self-energies. 
The entry in the pseudogap in the nodal direction of all other methods differs from the DiagMC result, as can be seen on the lower left panel where the green and blue symbols differ. 
However, while AFM-DMFT gives a better \emph{qualitative} agreement with DiagMC at the nodal point, PM-DMFT gives a better \emph{quantitative} agreement overall, that is for the whole frequency range. 

The self-energy obtained when only doing the local self-energy substitution on an \emph{ansatz} TPSC calculation is also plotted (solid red line).
It is clearly not sufficient to get an accurate total self-energy, since the low frequency part of the self-energy is not changed appreciably by this substitution. 
The use of both the DMFT double occupancy and self-energy is important to get accurate results for this model.


\section{Summary of the main results}
\label{section:summary_main_results}

As was seen in detail in \secref{section:results}, the TPSC and TPSC+DMFT methods give qualitatively similar results for the model studied here, namely the $2D$ Hubbard model on a square lattice. 
The method that better reproduces benchmark DiagMC quantities depends on the observable of interest. We present here a short summary of those results and the differences between the  methods.

The double occupancy is generally underestimated in TPSC and overestimated in TPSC+DMFT, as seen on Figs.~\ref{fig:comp_D_beta_demi_rempli} to~\ref{fig:comp_D_U_n_simkovic} of \secref{section:results_double_occ}.
The best qualitative agreement with benchmark DiagMC is from TPSC, which is often also quantitatively better than TPSC+DMFT. 

The spin susceptibility from both methods, shown in Figs.~\ref{fig:comp_chispmax_U_T} and~\ref{fig:comp_chisp_Q_path} of \secref{section:results_spin_susceptibility}, is close to benchmarks in the regime $U\lesssim 5$, with TPSC+DMFT having a better quantitative agreement with DiagMC close to half filling and low temperature. 
Neither TPSC method can reproduce the decrease in spin susceptibility caused by the Heisenberg-Mott physics for stronger interaction strength. 

The spin correlation length extracted at half filling is closer to benchmarks for TPSC+DMFT, as shown in Fig.~\ref{fig:comp_xisp_beta} of \secref{section:results_magnetic_corr_length}. 
Constraining the DMFT calculation to a paramagnetic solution leads to a better magnetic correlation length than allowing it to stabilize an antiferromagnetic long-range order.

The charge susceptibility from both TPSC methods is in good qualitative agreement with benchmarks for $U\lesssim 5$, as seen in Figs.~\ref{fig:comp_chichmax_U_T} and~\ref{fig:comp_chich_Q_path} of \secref{section:results_charge_susceptibility}. 
TPSC+DMFT gives slightly more accurate results for $2<U<5$. 
In particular, the temperature dependence of the charge susceptibility changes sign at $U\approx 4$ in TPSC, which is not the case in DiagMC or TPSC+DMFT.

In Fig.~\ref{fig:comp_self_energy_demi_rempli_U_2} of \secref{section:results_self_energy}, we show that at half filling, the momentum-resolved self-energy is more accurate in TPSC+DMFT, when using both the double occupancy and local part of the self-energy from DMFT. 
The use of antiferromagnetic or paramagnetic DMFT below the DMFT Néel temperature leads to a qualitatively different self-energy. 
The antiferromagnetic DMFT solution below the Néel temperature is in better qualitative agreement for the small Matsubara frequencies that reflect pseudogap physics. 
At higher Matsubara frequencies, the paramagnetic DMFT performs better.
Both the double occupancy and local part of the self-energy from DMFT are important in order to get an accurate total self-energy.


\section{Discussion}
\label{section:discussion}

When are TPSC+DMFT results in better agreement with benchmarks than TPSC and is there a way to estimate the accuracy of the results when no benchmarks are available? We propose answers to these questions here.  

\subsection{Why TPSC+DMFT?}
\label{section:discussion_compensation_errors}

\begin{figure}[htb]
\centering
\includegraphics[width=\figwidth]{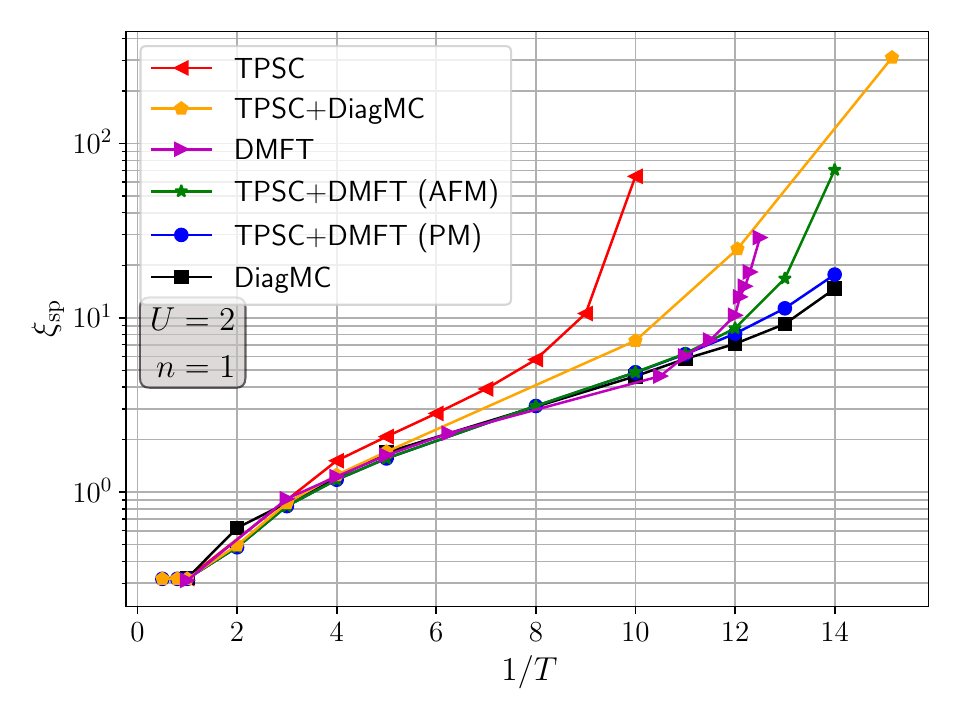}
\caption{Magnetic correlation length of the model at half filling, as already displayed in \figref{fig:comp_xisp_beta}.
The results obtained if we use for double occupancy the exact result from DiagMC instead of DMFT are also displayed in orange.
The result is then less accurate.
TPSC+DMFT in the paramagnetic state gives the best agreement with DiagMC for the antiferromagnetic correlation length.}
\label{fig:comp_xisp_beta_QMC}
\end{figure}

We have seen that combining TPSC with DMFT gives better quantitative agreement with benchmark DiagMC results for the spin susceptibility and the magnetic correlation length. 
This can be understood as a consequence of cancellation of errors. 
Indeed, as shown in Figs.~\ref{fig:comp_D_beta_demi_rempli}, \ref{fig:comp_D_T_U_simkovic_n_1.000}, and \ref{fig:comp_D_T_U_simkovic_n_0.875} of \secref{section:results_double_occ}, TPSC tends to underestimate double occupancy, while PM-DMFT and AFM-DMFT tend to overestimate it. 
Since TPSC overestimates spin fluctuations, using the larger DMFT double occupancy in the local-spin sum rule decreases $U_{\mathrm{sp}}$, hence giving better quantitative agreement with exact results, as shown in \figref{fig:comp_chispmax_U_n} and \figref{fig:comp_xisp_beta}. 
The improvement is also seen, to a lesser extent, with the charge fluctuations shown in \figref{fig:comp_chichmax_U_T}.
In some sense, there is a cancellation of errors between both methods when used together.
We have checked in a few cases that using the exact DiagMC result for double occupancy instead of that from DMFT gives less accurate results for the spin fluctuations at half filling, as seen in \figref{fig:comp_xisp_beta_QMC}.
Indeed, the double occupancy from DiagMC gives a more accurate spin correlation length (yellow pentagons) compared to TPSC alone (red triangles), but it is still too large.
The double occupancy from PM-DMFT, which is larger than the DiagMC value, gives a more accurate spin correlation length (blue circles).

\begin{figure}[hbt]
\centering
\includegraphics[width=\figwidth]{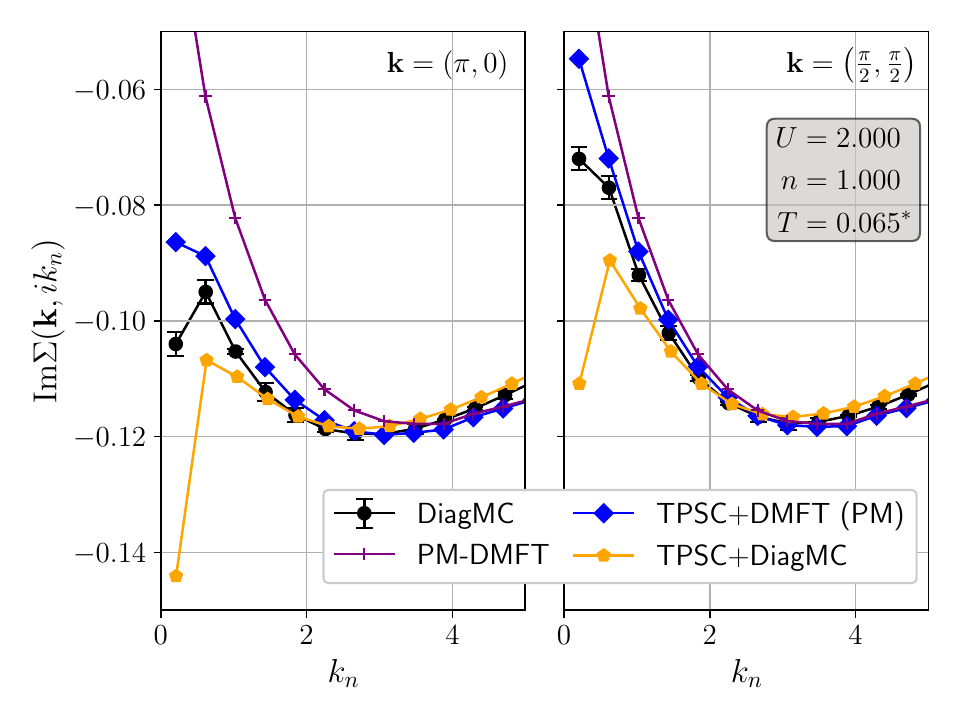}
\caption{Imaginary part of the self-energy as a function of Matsubara frequency $k_n$ at the antinode (left panel) and at the node (right panel). 
All the calculations were done at a temperature of $T=0.065$, except for TPSC-MC+$\Sigma^{\mathrm{loc}}$ which was done at $T=0.067$.
The local paramagnetic DMFT self-energy is also displayed in purple.
Error bars on all methods but DiagMC are smaller than the markers and not shown.
All DiagMC self-energy and double occupancy-data is taken from~\cite{Schafer_2021}.}
\label{fig:comp_self_energy_demi_rempli_U_2_QMC}
\end{figure}

What about the self energy? At high temperatures, we have verified that using either local self-energy from DMFT or DiagMC is a reasonable choice (not shown), as expected since the effect of the non-local fluctuations is small in this regime.
At $T=0.065$ (below the DMFT Néel temperature), the inclusion of the local observables from PM-DMFT, namely double occupancy and the local self-energy, gives the best quantitative agreement with exact results when combined with the non-local part of the self-energy from TPSC.
This is shown in \figref{fig:comp_self_energy_demi_rempli_U_2_QMC} by the blue diamonds.
The local DMFT self-energy alone is not sufficient at this temperature, as can be seen from the purple solid line.
It is more reasonable at high temperature~\cite{Schafer_2021}, where the local effects dominate. 
We have shown in \figref{fig:comp_self_energy_demi_rempli_U_2} that, for this model, the use of only the local part the the self-energy from DMFT is not enough. The effect of the DMFT double occupancy on the susceptibilities is also crucial to get an accurate total self-energy.
For a more accurate indication of the entrance in the pseudogap regime (indicated by the positive slope at the lowest frequencies), using  the exact double occupancy and local part of the self-energy from DiagMC (yellow pentagons in \figref{fig:comp_self_energy_demi_rempli_U_2_QMC}) or AFM-DMFT (green squares in \figref{fig:comp_self_energy_demi_rempli_U_2}) is preferable.

\begin{figure}[htb]
\centering
\includegraphics[width=\figwidth]{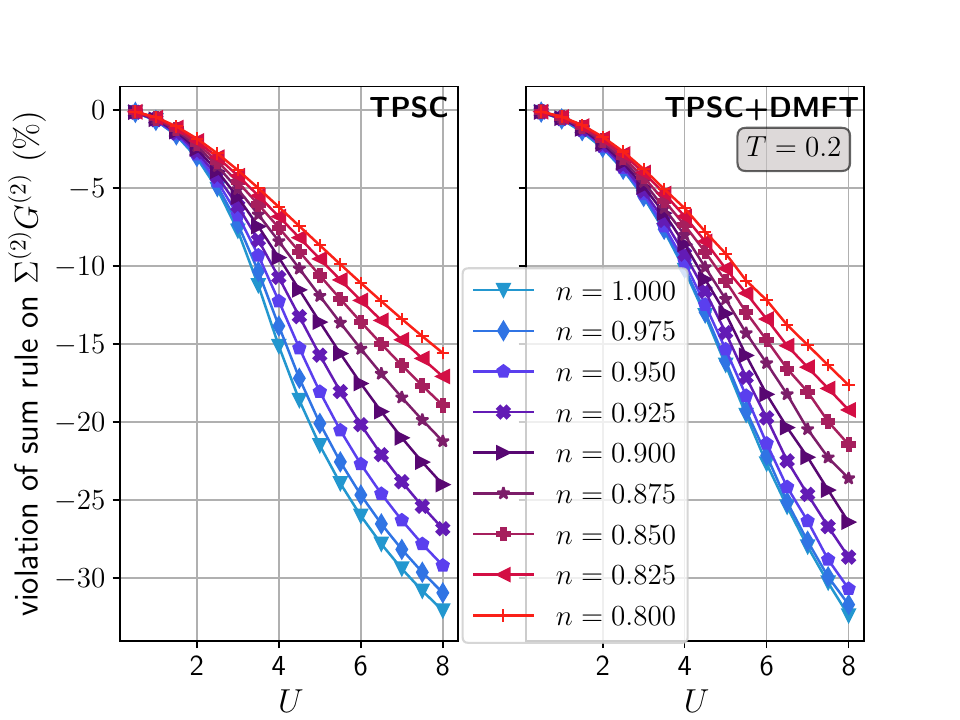}
\caption{Relative violation of the accuracy check \eref{eq:accuracy_check_Sig2G2} for both TPSC methods, corresponding to the data displayed in \figref{fig:comp_chispmax_U_n}, at $T=0.2$.}
\label{fig:comp_sumrule_TrS2G2_U_n}
\end{figure}

\begin{figure}[htb]
\centering
\includegraphics[width=\figwidth]{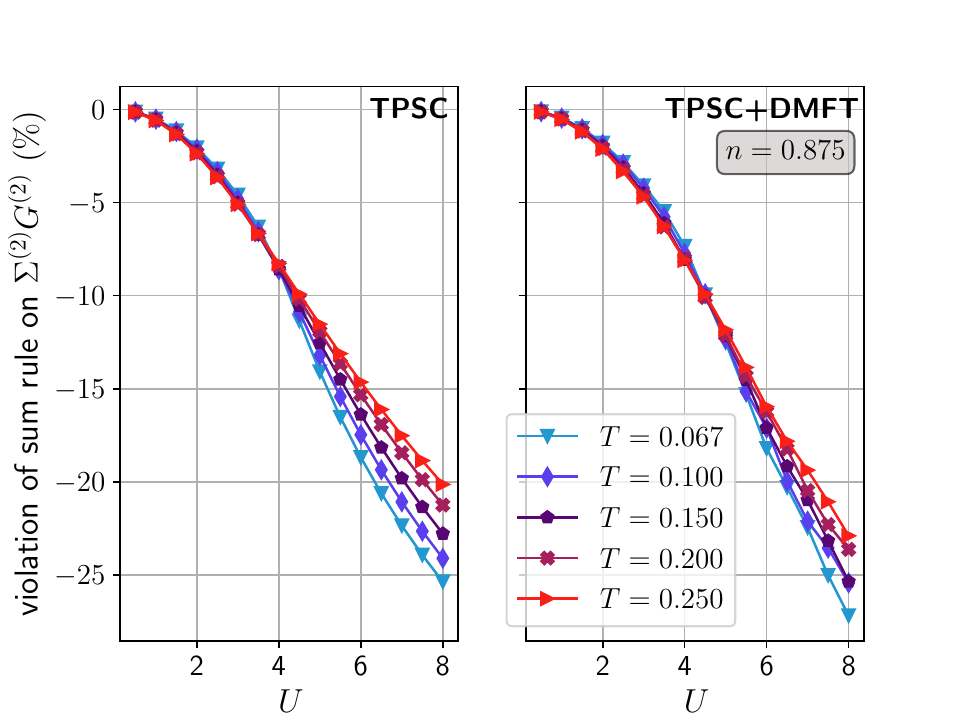}
\caption{Relative violation of the accuracy check Eq.~\ref{eq:accuracy_check_Sig2G2} for both TPSC methods, corresponding to the data displayed on Figs.~\ref{fig:comp_chispmax_U_T}~and~\ref{fig:comp_chichmax_U_T}, at $n=0.875$.}
\label{fig:comp_sumrule_TrS2G2_U_T}
\end{figure}

\subsection{Internal accuracy check}
\label{section:discussion_accuracy_check}

It is important to know if there is a way to estimate the accuracy of TPSC+DMFT in the absence of benchmarks. 
Such a method was proposed  for TPSC in Ref.~\cite{Vilk_1997} and explained in \eref{eq:accuracy_check_Sig2G2}. 
For $T=0.2$ and various fillings, \figref{fig:comp_sumrule_TrS2G2_U_n} shows as a function of $U$ the violation of the accuracy check corresponding to the susceptibility presented in \figref{fig:comp_chispmax_U_n}.
Note that the sum rule is computed before the local part of the self-energy from TPSC+DMFT is replaced by the PM-DMFT self-energy.
The general behavior is the same for TPSC and TPSC+DMFT. For $U\lesssim 5$, the violation of the sum rule is below $10\%$, and grows with $U$. 
This is consistent with what we observe for the spin susceptibility \figref{fig:comp_chispmax_U_n}. Indeed, this interaction strength $U$ corresponds to the point where the agreement with DiagMC results deteriorates. 
In addition, the violation of the sum rule is larger close to half filling, which is consistent with the spin susceptibility of both methods. 
We also see that violation of the sum rule is larger close to half filling for TPSC than for TPSC+DMFT, which is also consistent with the spin susceptibility. 

For $n=0.875$, and various temperatures, \figref{fig:comp_sumrule_TrS2G2_U_T} shows as a function of $U$ the violation of the accuracy check corresponding to the spin and charge susceptibilities presented in \figref{fig:comp_chispmax_U_T} and~\figref{fig:comp_chichmax_U_T}. The temperature dependence of this discrepancy is negligible for the interaction strengths for which TPSC is applicable, which is consistent with the corresponding susceptibility results. For large interaction strength $U\gtrsim 5$, the violation of the sum rule by TPSC+DMFT is larger than for TPSC. Neither method is reliable for large interaction strengths.

We therefore learn that the internal accuracy check developed for TPSC is also applicable for TPSC+DMFT for the model and regimes of parameters studied in this work, and is consistent with what the benchmarks suggest.


\section{Conclusion}

Our main results have been summarized in~\secref{section:summary_main_results}. 
(Data for the figures are available as Supplementary Material~\footnote{See Supplemental Material for the raw data and python codes used to generate the figures in this manuscript.}).
TPSC is a method that treats non-local correlations in a way that satisfies sum rules, conservation laws and the Mermin-Wagner theorem.
Single-site DMFT, on the other hand, emphasizes the role of local quantum fluctuations.
We have shown that combining TPSC with the DMFT double occupancy \emph{and} local self-energy  leads to a method, TPSC+DMFT, that can improve the agreement with benchmark DiagMC results in the weak to intermediate interaction range, $U\lesssim 5$. 
Even for $U\approx 5$, spin and charge fluctuations away from half filling can be qualitatively correct (\figref{fig:comp_chisp_Q_path} and \figref{fig:comp_chich_Q_path}).
TPSC+DMFT has a slightly widened regime of applicability near half filling and low temperature compared to TPSC. 
In agreement with previous benchmarking efforts~\cite{Vilk:1994,Veilleux:1995,Vilk:1996,Vilk_1997,Moukouri:2000,Kyung_2001,TremblayMancini:2011,Schafer_2021}, our current work shows that the TPSC approach is quantitatively accurate far from the renormalized classical regime and is qualitatively correct in general, overestimating however the temperature at which the renormalized classical regime begins.

The internal accuracy check devised for the original TPSC also works for TPSC+DMFT.
The violation of the internal accuracy check remains correlated with the deviation from the exact DiagMC results for the model and parameter regimes studied. 
This suggests that this accuracy check might be useful to assess reliability of the results in regimes where  no exact result is available.

In addition to providing a method that gives quantitatively accurate results in many regimes, the most important contribution of our work is to open the road to systematic multiband generalizations of TPSC that do not need the new \emph{ansatz} introduced in Refs.~\cite{Miyahara_2013,Ogura_Kuroki_2015,Zantout_2019,Zantout_2021}.
Steps in that direction have been taken~\cite{Zantout_2022}.
Multiband generalizations would be useful in realistic electronic-structure calculations.
Most contemporary codes include DMFT modules.
Adding the TPSC+DMFT option could be done at negligible computational cost and would allow inclusion of spin fluctuations to achieve even more realistic electronic structure calculations.


\section*{Acknowledgments}
We are grateful to S. Backes, D. Lessnich, A. Razpopov, R. Valent\'i and K. Zantout for useful discussions and for sharing the preprint of their related research~\cite{Zantout_2022}. We are especially grateful to the authors of Ref.~\cite{Simkovic_2022_benchmark}, F. \v{S}imkovic, R. Rossi and M. Ferrero for sharing the DiagMC results that we used as benchmarks.
We are also grateful to T. Sch\"afer and the authors of Ref.~\cite{Schafer_2021} for making their benchmarks publicly available.
This work has been supported by the Natural Sciences and Engineering Research Council of Canada (NSERC) under grant RGPIN-2019-05312, by an Excellence Scolarship (N. M.) from Hydro-Qu\'ebec, by a Vanier Scholarship (C. G.-N.) from NSERC and by the Canada First Research Excellence Fund. 
Simulations were performed on computers provided by the Canadian Foundation for Innovation, the Minist\`ere de l'\'Education des Loisirs et du Sport (Qu\'ebec), Calcul Qu\'ebec, and the Digital Reseach Alliance of Canada.

\FloatBarrier
\bibliography{bibliography}

\end{document}